%% file: main.tex
\documentclass[sigplan, screen]{acmart}

\input{config/packages}
\input{config/macros}

\AtBeginDocument{%
  \providecommand\BibTeX{{%
    \normalfont B\kern-0.5em{\scshape i\kern-0.25em b}\kern-0.8em\TeX}}}

\setcopyright{none}
\renewcommand\footnotetextcopyrightpermission[1]{}
\acmConference[Preprint]{Preprint}{arXiv}{2026}
\begin{document}

\title
[\sysname{}: Transparent Thread-Level Offloading on Transient Cloud Resources Using MPKs]
{\huge \sysname: Transparent Thread-Level Offloading\\on Transient Cloud Resources Using MPKs}

\include{config/authors}
\input{content/abstract}

\maketitle

\input{content/intro}
\input{content/background}
\input{content/overview}
\input{content/coherence}
\input{content/implementation}
\input{content/evaluation}
\input{content/discussion}
\input{content/related}
\input{content/conclusion}

\newpage

\bibliographystyle{ACM-Reference-Format}
\bibliography{references}

\end{document}

%% file: config/packages.tex
\usepackage{hyperref}
\usepackage{listings}
\usepackage{enumitem}
\usepackage{xspace}
\usepackage[scaled=0.9]{beramono}
\microtypecontext{spacing=nonfrench}

\makeatletter
\lst@Key{countblanklines}{true}[t]%
{\lstKV@SetIf{#1}\lst@ifcountblanklines}

\lst@AddToHook{OnEmptyLine}{%
	\lst@ifnumberblanklines\else%
	\lst@ifcountblanklines\else%
	\advance\c@lstnumber-\@ne\relax%
	\fi%
	\fi}
\makeatother

%% file: config/macros.tex
\newcommand{\sysname}{ThreadShift\xspace}

\newcommand{\us}{µs\xspace}
\renewcommand{\t}[1]{\texttt{\footnotesize #1}}

\newenvironment{myitemize}{\begin{list}{\labelitemi}{%
\setlength{\topsep}{0.5pt plus 0pt minus 0pt}%
\setlength{\itemsep}{0pt plus 0pt minus 0pt}%
\setlength{\parsep}{0pt plus 0pt minus 0pt}%
\setlength{\parskip}{0pt plus 0pt minus 0pt}%
}}{\end{list}}

%% file: config/authors.tex
\author{Antoine Murat}
\affiliation[obeypunctuation=true]{%
  \institution{EPFL}, \country{Switzerland}%
}
\author{Clément Burgelin}
\affiliation[obeypunctuation=true]{%
  \institution{EPFL}, \country{Switzerland}%
}
\author{Rachid Guerraoui}
\affiliation[obeypunctuation=true]{%
  \institution{EPFL}, \country{Switzerland}%
}

\makeatletter
\renewcommand{\@authorfont}{\normalsize\sffamily}
\renewcommand{\@affiliationfont}{\footnotesize\normalfont}
\makeatother

%% file: content/abstract.tex
\begin{abstract}
Transient cloud resources offer significant cost savings, but their unpredictability makes them hard to use for applications that cannot be safely restarted after reclamation.
Existing approaches require application changes or rely on coarse-grained checkpointing, whose cost limits its benefits.

We present \emph{\textbf{\sysname}}, a system that transparently offloads compute-intensive threads of unmodified Linux applications to cheap transient resources while preserving correctness under reclamation. By operating at thread granularity, \sysname enables fast, fine-grained checkpointing and offloads only the threads that benefit from transient execution.


The main challenge is to checkpoint individual threads despite cross-thread dependencies. \sysname does so by efficiently tracking memory dependencies, identifying per-thread memory writes, and maintaining coherence across machines. Three novel uses of Memory Protection Keys (MPKs) make these mechanisms efficient, enabling fast incremental checkpoints without pausing the entire application.

We implement \sysname on x86-64 Linux and evaluate it on workloads including machine-learning inference, cryptographic tasks, and in-memory data processing. \sysname offloads threads in as little as 164\,\us, performs checkpointing up to three orders of magnitude faster than CRIU, and reduces deployment cost by up to 56\% on commodity clouds.
\end{abstract}

%% file: content/intro.tex
\section{Introduction}\label{sec:intro}

Transient cloud resources offer significant cost savings by exposing temporarily idle capacity at a discount (e.g., spot VMs~\cite{use-azure-spot, gcp-spot, ec2-spot}). However, these resources can be reclaimed at any time, making them difficult to use for applications that cannot be safely restarted after reclamation. In such applications, execution may already have produced externally visible effects (e.g., sent requests), and restarting would lead to duplicated or inconsistent outcomes. Even distributed applications designed to tolerate failures can struggle with the higher frequency and correlation of such reclamations~\cite{t-basir}.

A natural approach to tolerate reclamations is checkpointing, i.e., periodically saving application state so execution can resume after interruption~\cite{survey-rollback}. In practice, however, most applications lack built-in checkpointing and must rely on general-purpose systems such as CRIU~\cite{criu}. These systems operate at coarse granularity (e.g., processes or containers), incurring high overhead and often pausing the entire application. As a result, applications that require frequent checkpoints cannot effectively benefit from transient resources.

This paper asks: \emph{can unmodified applications leverage transient resources without costly coarse-grained checkpointing?}

We present \emph{\sysname}, a system that transparently offloads compute-intensive threads of unmodified Linux applications to transient resources with low overhead while preserving correctness under reclamation. Here, correctness means that clients cannot distinguish execution on unreliable from reliable resources. The key idea is to execute only the parts of the application that can safely benefit from cheap reclaimable resources on transient machines, while keeping the rest on a weak, inexpensive stable machine that also maintains checkpoints and handles external effects.

\sysname operates at thread granularity. This is critical: it enables fast, fine-grained checkpointing, avoids blocking the entire application, and allows threads to run on stable or transient resources based on behavior. As a result, compute-heavy threads can be offloaded aggressively, while IO-bound or latency-sensitive threads can remain on stable machines.

Making this design practical is challenging. Threads in a process share memory and interact in subtle ways, so checkpointing one thread in isolation is unsafe: its state may depend on memory modified by others. Efficiently capturing consistent checkpoints therefore requires tracking inter-thread dependencies, identifying per-thread memory modifications, and maintaining a coherent view of memory across machines. Naively implementing these mechanisms would incur high overhead due to costly instrumentation, frequent page protection, and global synchronization.

\sysname addresses these challenges using three novel applications of Memory Protection Keys (MPKs). Rather than using MPKs for security, \sysname uses per-thread permissions to cheaply detect cross-thread dependencies on first access, track memory modifications without freezing the full application, and avoid costly per-page protection changes. This allows \sysname to capture thread-level checkpoints and maintain memory coherence with low overhead.

We implement \sysname on x86-64 Linux and evaluate it on a range of workloads, including machine-learning inference, cryptographic tasks, and in-memory data processing. Our results show that \sysname can offload threads in as little as 164\,\us, perform checkpointing up to three orders of magnitude faster than CRIU, and reduce deployment cost by up to 56\% on commodity clouds.

In summary, our contributions are the following:
\begin{myitemize}
    \item \sysname, a system that offloads threads of unmodified Linux applications to transient resources while preserving correctness under reclamation.
    \item An MPK-based mechanism for tracking inter-thread memory dependencies with low overhead.
    \item An MPK-based mechanism for efficiently identifying per-thread memory modifications.
    \item An MPK-based mechanism for reducing the overhead of software memory coherence.
    \item An open-source prototype of \sysname, available at \href{https://github.com/AntoineMurat/threadshift}{https://github.com/AntoineMurat/threadshift}.
    \item A thorough evaluation of \sysname demonstrating substantial reductions in checkpointing overhead and deployment cost across diverse workloads.
\end{myitemize}

%% file: content/background.tex
\section{Background} \label{sec:background}

After some background on checkpointing~(\S\ref{sec:background:checkpointing}), thread migration~(\S\ref{sec:background:migration}), and Memory Protection Keys (MPKs)~(\S\ref{sec:background:mpk}), we present our system model~(\S\ref{sec:background:model}) and target applications~(\S\ref{sec:background:target}).

\subsection{Application Checkpointing}\label{sec:background:checkpointing}

Deploying applications on transient resources has led to numerous proposals~\cite{
    DBLP:conf/eurosys/JiaSSRW16,
    bamboo,
    varuna,
    spot-mem-cache,
    elastic-datastore-spot,
    blend-spot-memstore,
    spotcheck,
    spoton,
    hotspot-server-hop}.
Many of these rely on Checkpoint/Restore (CR) schemes~\cite{survey-rollback}, which periodically save application state so execution can resume after a reclamation. \sysname follows this approach.

Some applications use checkpoints primarily to save work after failures~\cite{bamboo,varuna,spoton}, while others require stronger guarantees (e.g., avoiding contradictory replies) and therefore need timely checkpoints (e.g., before sending messages) to preserve external consistency~\cite{responsiv-repl-container, plover, remus}.%
\footnote{Critical events can also be logged for replay to avoid timely checkpoints, but capturing all non-deterministic events is challenging in practice~\cite{dthreads,responsiv-repl-container}.}
\sysname targets the latter setting and checkpoints applications upon each event that affects their external state.

\begin{figure}
 \includegraphics[width=\columnwidth]{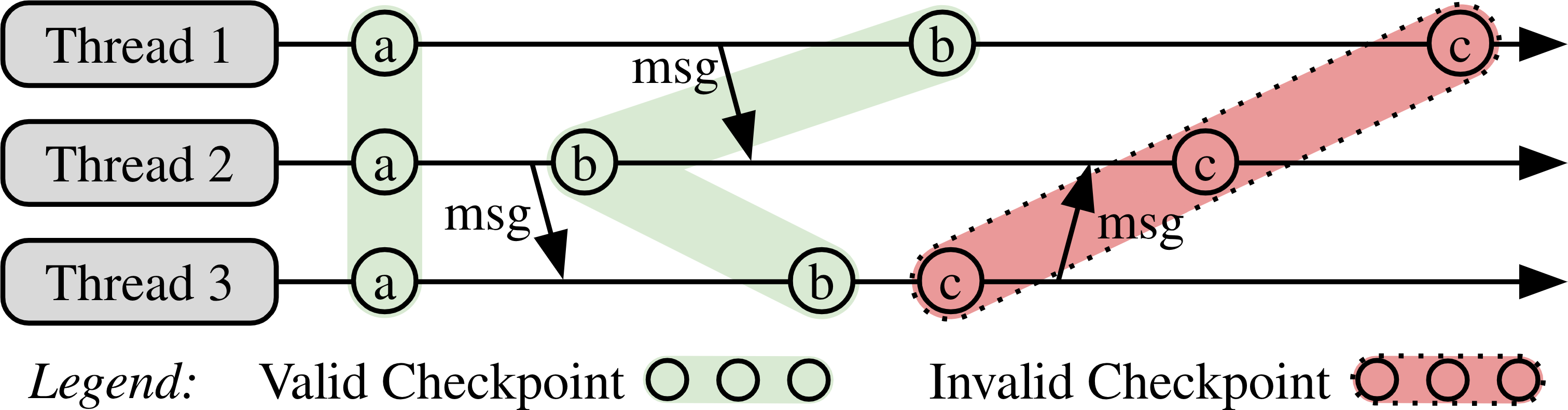}
 \caption{Application checkpoints composed of individual thread checkpoints must capture all dependencies. Checkpoint $c$ is invalid: it includes Thread 2 after receiving a message from Thread 3, but Thread 3 before sending it; restoring Thread 3 would result in the message being sent twice.}
 \label{fig:checkpoints}
\end{figure}

Application-agnostic CR mechanisms typically operate at coarse granularity, such as processes~\cite{criu} or VMs~\cite{plover, remus}. \sysname instead operates at a finer \emph{thread} granularity. This is challenging because thread-level checkpointing must track memory modifications and capture inter-thread dependencies. As illustrated in Figure~\ref{fig:checkpoints}, application checkpoints must include consistent views of interacting threads; otherwise, restoring them can duplicate or suppress effects~\cite{message-dependencies}.

\subsection{Thread Migration} \label{sec:background:migration}

Offloading a thread requires migrating its (1) CPU context (i.e., registers), (2) memory, and (3) kernel objects accessed via syscalls (e.g., files). CPU context is private, but memory and kernel objects are shared with other threads, requiring careful coherence control to maintain consistency across machines.
\sysname avoids synchronizing kernel objects by reloading threads before they execute syscalls~\cite{sprite}, so syscalls always run on the stable machine. Memory must still be kept consistent across machines, which \sysname handles using a page-level coherence protocol (Section~\ref{sec:mc}).

\subsection{Memory Protection Keys} \label{sec:background:mpk}

\begin{figure}
 \includegraphics[width=\columnwidth]{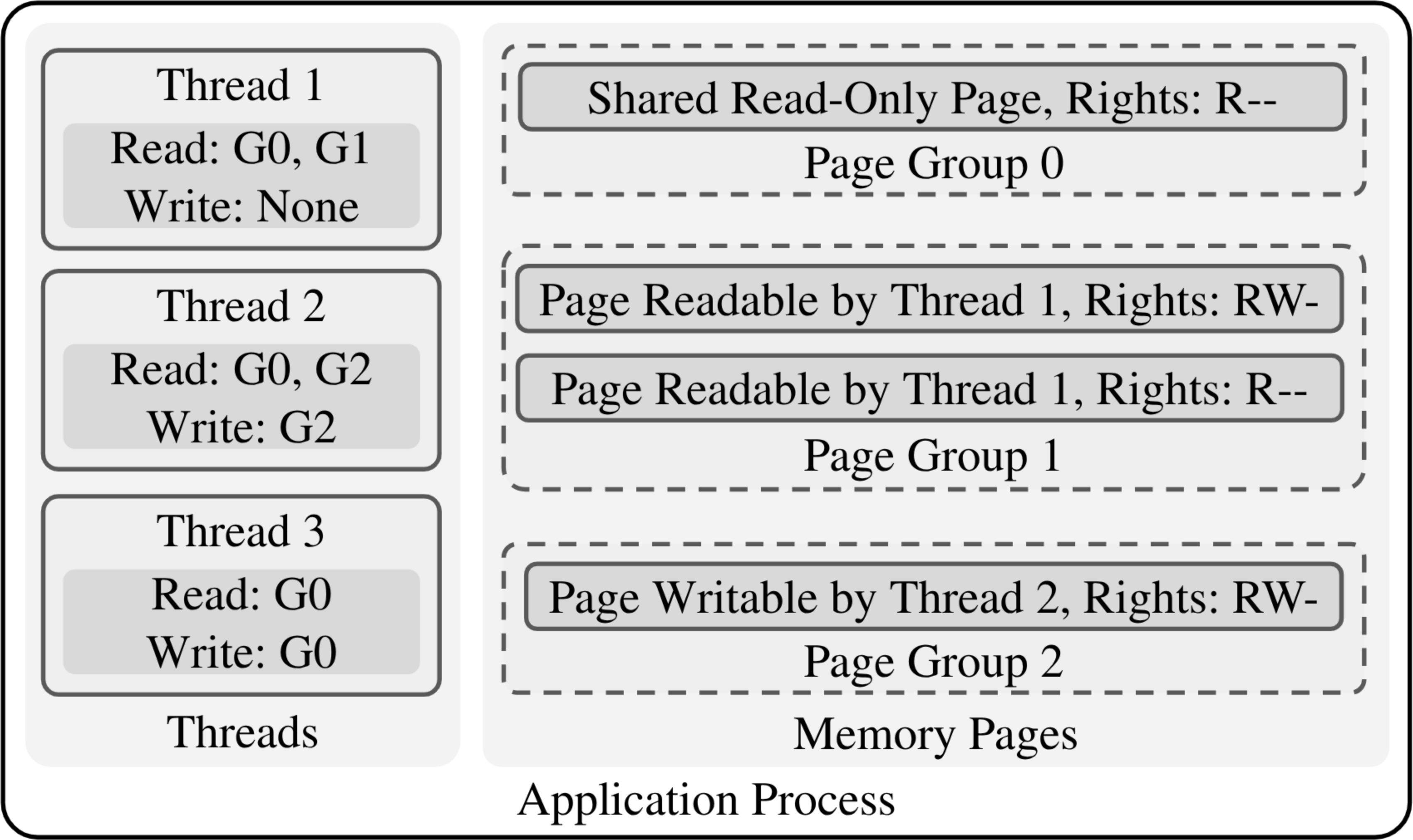}
 \caption{Example of MPKs. Thread 1 can read groups 0 and 1 but write to none due to its thread-local rights. Thread 2 can read groups 0 and 2 and write to group 2. Thread 3 can read group 0 but not write to it as the page is read-only.}
 \label{fig:pku}
\end{figure}

Memory Protection Keys (MPKs) are a hardware feature that allows applications to control memory access on a per-thread basis, beyond traditional page protections. MPKs partition memory into page groups, whose access rights can be independently controlled by each thread (Figure~\ref{fig:pku}).

Access permissions are governed by a per-core register called \emph{PKRU}, which can be updated from userspace. These updates are fast and do not trigger TLB invalidations. When a thread accesses a page, hardware checks both traditional permissions and MPKs; if either disallows the access, a fault is raised which can be handled in userspace.

\sysname leverages MPKs to track inter-thread dependencies, identify per-thread memory modifications, and control memory access during distributed execution (Section~\ref{sec:mc}).

\begin{figure*}[t!]
\centering
\includegraphics[width=\textwidth]{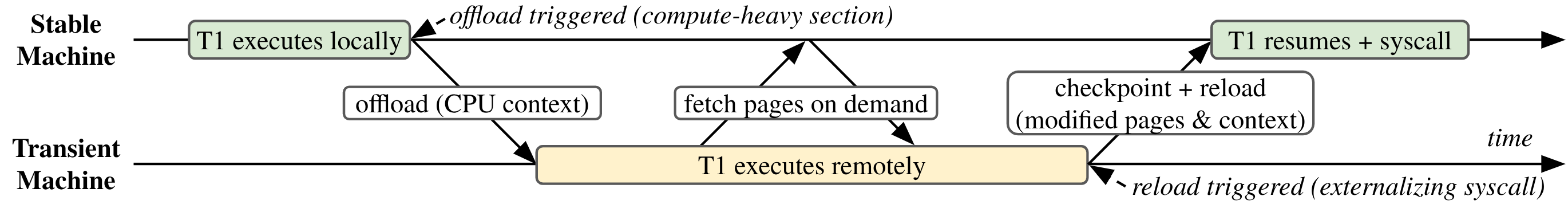}
\caption{Running example. The thread $T_1$ executes locally until it enters a compute-heavy section and is offloaded to the transient machine. It then executes remotely while fetching pages on demand. Before performing an externalizing syscall, it checkpoints its modified state and reloads on the stable machine, where execution resumes and the syscall completes.}
\label{fig:overview:timeline}
\end{figure*}

\subsection{System Model} \label{sec:background:model}

\sysname allows unmodified applications to run on transient resources without increasing their unreliability beyond that of an on-demand deployment. The system assumes that the stable component does not fail; making this component fault-tolerant is orthogonal to our work.

Transient resources may be reclaimed at any time, with or without prior notification. \sysname does not rely on such notifications for correctness: it treats reclamations as failures and recovers execution from checkpoints. When available, notifications are used opportunistically to migrate threads back to the stable machine, but \sysname does not assume sufficient notice, as notifications may be best-effort~\cite{use-azure-spot, gcp-spot} and may not leave enough time to complete state migration~\cite{spotweb, spoton}. If migration does not finish in time, \sysname simply recovers from the latest checkpoint.
More generally, \sysname makes no synchrony assumptions and treats unresponsive transient resources as failed.

Finally, \sysname assumes low-cost data transfers inside a data center, as is common on commodity cloud platforms.

\subsection{Target Applications} \label{sec:background:target}

\sysname targets applications that cannot be restarted upon reclamation without violating correctness (e.g., by sending duplicate responses or dropping requests), and that contain compute-intensive sections. \sysname provides benefits even when such sections last less than a millisecond. This includes workloads such as machine-learning inference (e.g., image classification and speech-to-text~\cite{whisper-cpp, resnet50, pytorch}), scientific and financial computations (e.g., simulations and option pricing~\cite{onemkl, openfoam, lammps}), and cryptographic applications (e.g., distributed ledgers and verifiable services~\cite{secp256k1,zero-knowledge-blockchain,homomorphic-encrypt-survey,argon2}).

\sysname can also batch requests to aggregate short compute-intensive tasks into larger ones, increasing the amount of work that can be offloaded and amortizing migration and checkpointing overheads (Section~\ref{sec:para:network}). In contrast, it is not well suited for IO-intensive applications or workloads that rely heavily on specialized hardware (e.g., ML training).

\sysname is relevant to both single-node applications that cannot tolerate reclamations and distributed applications that are vulnerable to frequent or correlated failures~\cite{t-basir}.

%% file: content/overview.tex
\section{\sysname in a Nutshell}\label{sec:overview}

This section gives a top-down view of \sysname. We first walk through a simple execution example that illustrates its main principles and challenges (\S\ref{sec:overview:example}). We then summarize the architecture that realizes this execution model (\S\ref{sec:overview:arch}) and describe how \sysname accelerates common code patterns through para-offloading (\S\ref{sec:para}). We defer the detailed coherence protocol and MPK-based mechanisms to Section~\ref{sec:mc}.

\subsection{Running Example and Principles}\label{sec:overview:example}

Consider a multi-threaded server with two worker threads, $T_1$ and $T_2$, that handle client requests end to end: each thread receives a request, enters a compute-intensive processing phase, and then sends a reply. The compute-intensive phase is the part that \sysname offloads. Most of the time, $T_1$ and $T_2$ process requests independently, but they occasionally interact through shared memory. For example, if $T_2$ becomes idle, it may steal work from $T_1$'s queue. Such interactions create cross-thread dependencies that must be accounted for.

The server also includes a third thread, $T_C$, responsible for control-plane tasks such as applying configuration updates. Because $T_C$ is not compute-intensive and not on the request-processing fast path, it remains on the stable machine.

\paragraph{Offloading.}
When $T_1$ enters its processing phase, \sysname offloads it to a transient machine (Figure~\ref{fig:overview:timeline}). Concretely, the stable machine snapshots $T_1$'s execution context, i.e., its thread-private registers, and sends them to a worker. This is analogous to a context switch, except that the saved context is reloaded on another machine rather than on another core.

Execution then resumes remotely. At first, the worker may miss pages that $T_1$ needs, such as its stack, code, or input data. As $T_1$ touches them, the worker fetches them on demand. While running remotely, $T_1$ reads memory, performs compute-intensive work, and writes its output.

\paragraph{Externalization.}
Suppose now that $T_1$ is ready to reply to the client. The reply is about to become externally visible, so \sysname must ensure that the state leading to that reply is recoverable. The worker therefore checkpoints $T_1$'s execution context together with the memory it modified, sends this checkpoint to the stable machine, and requests a reload. The stable machine applies the memory updates and resume execution locally by installing $T_1$'s latest execution context. The \t{send} syscall then executes on the stable machine.

This design is deliberate: \sysname offloads compute, but externalization occurs on the stable machine to ensure correctness despite reclamations. More generally, all system calls execute there, so kernel-managed state (e.g., sockets) remains local and need not be synchronized across machines.

\vspace{0.5em}
The same execution pattern applies to $T_2$. In the common case, $T_1$ and $T_2$ handle different requests and can therefore be offloaded and checkpointed independently. This keeps checkpointing local to the thread that is about to externalize its result, rather than forcing unrelated work to be checkpointed, as in conventional process-wide checkpointing.

\paragraph{Dependencies.}
Now consider a less independent execution. Suppose that $T_2$ runs out of local work and steals a request from $T_1$'s queue. At that point, $T_2$ has read memory written by $T_1$, so the two threads are no longer independent. If \sysname later checkpoints $T_2$, checkpointing $T_2$ alone would be unsafe: its state now depends on $T_1$. \sysname must therefore detect this interaction and checkpoint both threads consistently before exposing their results.

\paragraph{Reclamation.}
Finally, suppose that the transient machine hosting $T_1$ and $T_2$ is about to be reclaimed. If the warning arrives early enough, \sysname starts migrating the affected threads back to the stable machine. In our example, however, the warning is not sufficient to checkpoint both dependent threads in time. In that case, \sysname treats the reclamation as a failure and resumes execution on the stable machine from the latest checkpoints that were already safely stored there. This may cause some recomputation, but it preserves correctness, e.g., it never duplicates replies.

\paragraph{Challenges.}
The difficulty is not just to handle these cases correctly, but to do so cheaply enough that offloading remains worthwhile. In the example above, \sysname must detect when threads become dependent through shared memory. It must also identify which memory each checkpointed thread modified so that checkpointing remains fast. Finally, all machines must maintain a coherent view of memory.

Naively implementing these mechanisms would make thread-level offloading too expensive: tracking dependencies with instrumentation or page protections would incur high overhead, identifying modified memory could require pausing unrelated threads, and enforcing coherence could trigger many costly page-protection updates. Section~\ref{sec:mc} addresses these challenges with three MPK-based mechanisms that make dependency tracking, dirty tracking, and coherence enforcement efficient enough for fine-grained offloading.

\subsection{Architecture}\label{sec:overview:arch}

\begin{figure}
 \includegraphics[width=\columnwidth]{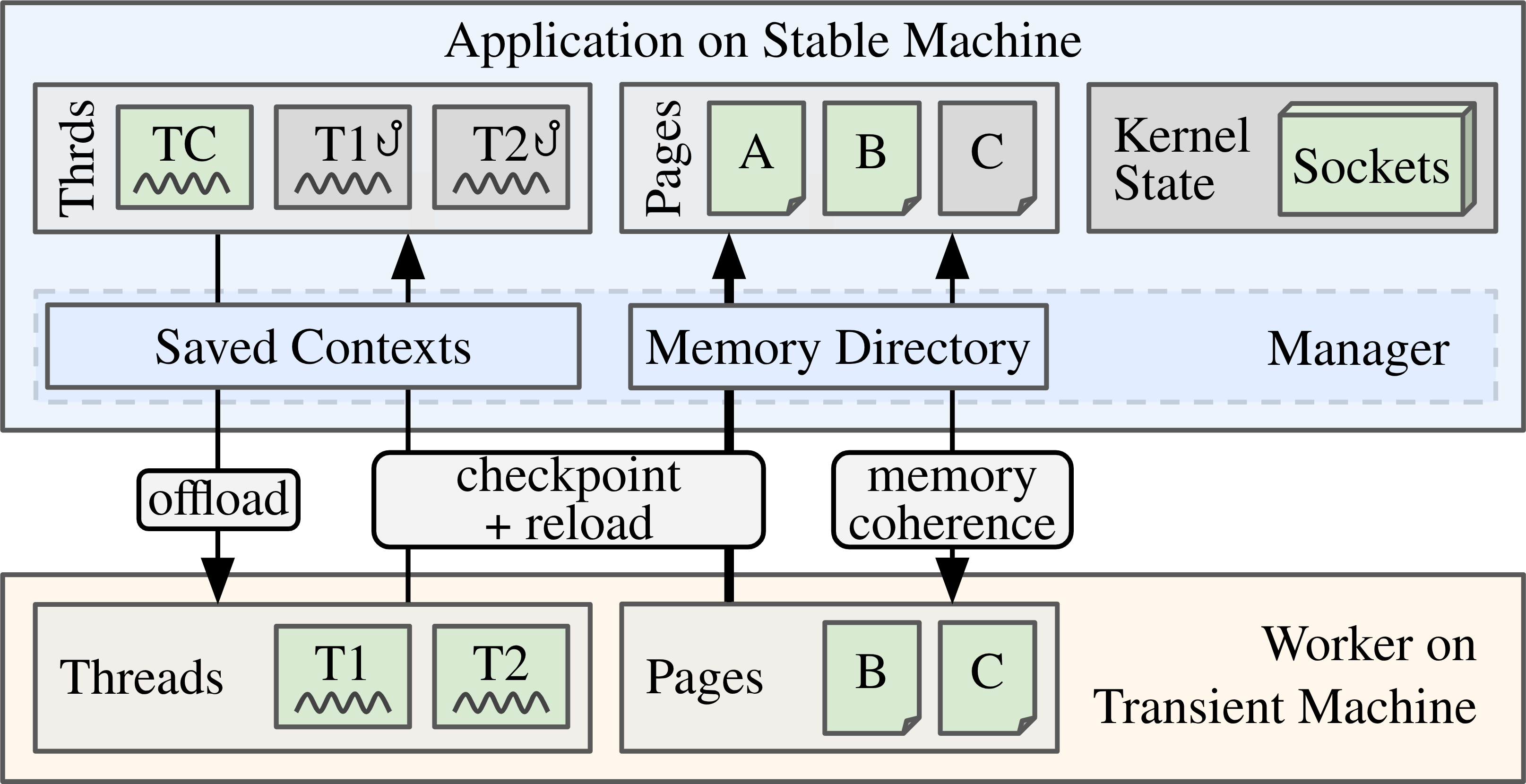}
\caption{
Overview of \sysname's architecture.
The application runs on a stable machine (top) with an injected manager that stores thread contexts and coherence metadata.
Threads are offloaded to workers on transient machines (bottom) and return upon checkpointing, while memory is kept coherent.
Kernel state remains on the stable machine.}
 \label{fig:arch}
\end{figure}

Figure~\ref{fig:arch} summarizes \sysname's architecture. An unmodified application runs on a stable machine with a lightweight \emph{manager} injected at startup. Compute-intensive threads are offloaded to \emph{workers} on transient machines. The stable machine handles checkpoints, externalization, and memory coherence, while workers execute offloaded computation.

\paragraph{Stable machine.}
The application's main process runs on the stable machine. At startup, \sysname injects a manager into this process via \t{LD\_PRELOAD}~\cite{glibc-ld}, allowing it to control execution flow and memory access. The manager identifies offload points, sends thread contexts to workers, stores checkpoints, and coordinates memory coherence across machines.

When a thread enters a compute-heavy section, the manager snapshots its CPU registers, keeps a local copy as a checkpoint, and sends the context to a worker. When a worker later returns an incremental checkpoint, the manager applies the reported memory updates and refreshes the saved execution context. If a worker requests a reload or becomes unavailable, the manager resumes the corresponding thread from the latest checkpoint on the stable machine.

\paragraph{Transient machines.}
Transient machines run workers, which are ordinary Linux processes with threads prepared to host offloads. When a worker receives an offload request, it installs the incoming execution context into one of its idle threads and resumes execution there.

Workers execute offloaded computation and synchronize memory with the stable machine, but they do not externalize results. When an offloaded thread reaches a system call, the worker pauses it, checkpoints its updated context and modified memory, and requests that execution continue on the stable machine. If the checkpoint depends on other threads, the worker atomically checkpoints them as well.

A worker may host multiple threads, and workers may span multiple transient machines. Collocating offloads reduces memory-coherence traffic, while distributing them might reduce cost and limits the impact of reclamations.

\paragraph{Why memory is the hard part.}
Within this architecture, registers are easy to move around and kernel-managed state remains local because all system calls execute on the stable machine. Memory is harder: it is shared across threads, accessed from multiple machines, and must remain recoverable under worker loss. \sysname therefore requires a tailored protocol that ensures not only coherence, but also that memory updates become visible only when they are backed by a safe checkpoint. Section~\ref{sec:mc} describes this protocol and the MPK-based mechanisms that make it efficient.

\subsection{Para-Offloading}\label{sec:para}

\sysname works with unmodified applications, but some common code patterns offload poorly. Para-offloading is an optional optimization layer that replaces selected library behavior with semantically equivalent, \sysname-friendly behavior. Importantly, it requires no changes to user code.

\paragraph{Network IO}\label{sec:para:network}
In the running example, worker threads follow a \t{recv}$\rightarrow$\t{process}$\rightarrow$\t{send} pattern. Directly offloading short requests is inefficient because every request triggers an offload and every reply forces a return to the stable machine. To amortize this overhead, \sysname intercepts socket APIs so that incoming streams can be forwarded directly to workers, while outgoing data is buffered there. When a worker checkpoints, it reports both the number of consumed input bytes and the buffered replies. The manager then advances its view of the input stream and flushes the buffered output. This batches short requests into larger offloads, increasing throughput at the cost of a modest increase in latency.

\paragraph{Memory allocation}\label{sec:para:memory}
Frequent memory allocation can hurt offloading because allocators may invoke syscalls and introduce unnecessary inter-thread synchronization due to false sharing. \sysname therefore replaces them with more offload-friendly allocators, typically trading a modest increase in memory use for more efficient offloading.

\paragraph{Synchronization}\label{sec:para:futexes}
More generally, synchronization hurts offloading when it relies on syscalls, as these trigger reloads on the stable machine. \sysname therefore replaces standard synchronization primitives (e.g., from libc) with specially handled variants that avoid such syscalls, reducing coordination overhead while preserving semantics.

%% file: content/coherence.tex
\section{\sysname's Memory Coherence}\label{sec:mc}

This section explains how \sysname keeps memory coherent across machines while ensuring that recovery from reclamation remains safe. We first present \sysname's coherence protocol at a high level (\S\ref{sec:mc:protocol}), and then describe the three MPK-based mechanisms that make it efficient: dependency tracking (\S\ref{sec:mc:deps}), per-thread dirty tracking (\S\ref{sec:mc:dirty}), and deferred protection (\S\ref{sec:mc:protect}).

\subsection{Protocol Overview}\label{sec:mc:protocol}

\sysname's Memory Coherence Protocol (TMCP) tracks which machine may read or write each page via a directory maintained by \sysname's manager on the stable machine. TMCP resembles MESI at a high level, but is tailored to thread offloading and abrupt worker loss. TMCP operates at page granularity, matching x86 permissions~\cite{intel2019developer}, but reduces transfer volume by comparing page chunks via hashes~\cite{plover}.


TMCP enforces two invariants. First, if a machine can write a page, no other machine may access it concurrently, and any readable copy reflects its latest value. Second, a worker's write becomes visible to other machines only after the threads that (transitively) produced it have been checkpointed, along with the memory they modified. Together, these invariants ensure coherence and recovery.

Recoverability matters because \sysname must survive worker loss. If a worker exposed writes before checkpointing the threads that produced them, a failure could make visible state revert to an earlier version upon recovery, leading to an inconsistent execution. TMCP prevents this by exposing worker writes only after they are backed by stable state.

TMCP runs on both the stable machine and workers, which maintain different state machines as their roles differ.

\subsubsection{Worker-side states}\label{sec:mc:protocol:workers}

\begin{table}[b]
\caption{Page states at a worker.}
\begin{tabular}{lcc}
State & Rights & Meaning \\ \hline
Modified  & RWX & Uncheckpointed writable copy. \\ \hline
Exclusive & RWX & Checkpointed writable copy. \\ \hline
Shared    & R-X & Read-only copy. \\ \hline
Invalid   & --- & No readable copy. \\ \hline
\end{tabular}
\label{table:state:workers}
\end{table}

Table~\ref{table:state:workers} summarizes worker-side states. The \t{Shared}, \t{Exclusive}, and \t{Invalid} states behave much like their MESI counterparts. The key difference is the meaning of \t{Modified}: a page in this state holds uncheckpointed writes that are not yet safe to expose.

Initially, worker pages are \t{Invalid}. Reading such a page sends a \t{Share} request to the manager and upgrades it to \t{Shared}, while writing sends a \t{Borrow} request and upgrades it to \t{Modified}. Writing a \t{Shared} page likewise issues a \t{Borrow} request, and writing an \t{Exclusive} page makes it \t{Modified}. Symmetrically, clean pages are downgraded upon \t{Unshare} and \t{Unborrow} requests to \t{Invalid} and \t{Shared}, respectively. \t{Modified} pages require extra care: before they can be downgraded or remain writable as \t{Exclusive}, the modifications they contain must first be made recoverable.

To do so, the worker atomically builds a checkpoint that covers the page: it records the page's changes, snapshots every thread that wrote to it since it entered \t{Modified}, and recursively includes any threads on which those writers depend, together with the other \t{Modified} pages they changed. Once this checkpoint is sent to the manager, all included pages become safe: pages the worker keeps writable move to \t{Exclusive}, while others are downgraded as requested.

Tracking dependencies is thus essential: without it, \sysname would need to conservatively checkpoint all offloads and modified data whenever a \t{Modified} page is downgraded. Section~\ref{sec:mc:deps} explains how to do so efficiently using MPKs.

\begin{figure*}
 \includegraphics[width=\textwidth]{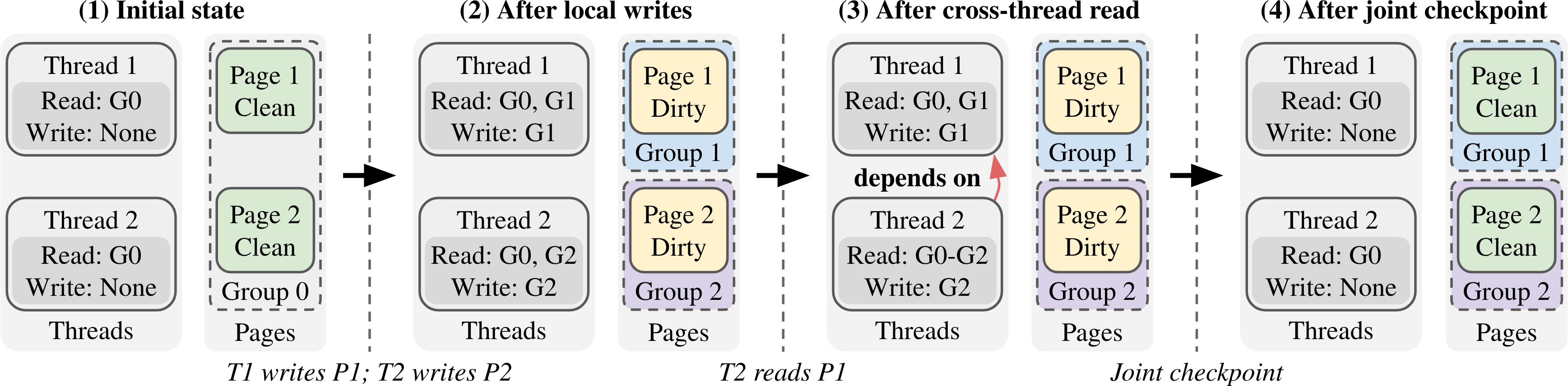}
\caption{MPK-based dependency tracking.
(1) Pages start in a shared group (\t{G0}) readable by all threads.
(2) A write moves a page to a dedicated group (\t{G1} or \t{G2}) that grants write access only to its writer.
(3) A later cross-thread read faults, revealing a dependency and extending permissions.
(4) When $T_2$ is checkpointed, the thread $T_1$ it depends on is also checkpointed; permissions and page states are reset, but groups are kept.}
 \label{fig:mpk-example}
\end{figure*}

\subsubsection{Stable-machine states}\label{sec:mc:protocol:parasite}

\begin{table}
\caption{Page states at the stable machine.}
\begin{tabular}{lcc}
State & Rights & Meaning \\ \hline
Exclusive & RWX & No worker has a readable copy. \\ \hline
Shared    & R-X & 1+ workers have a read-only copy. \\ \hline
Invalid   & --- & A worker has a writable copy. \\ \hline
\end{tabular}
\label{table:state:parasite}
\end{table}

The manager mediates all permission changes, serving both local accesses and worker requests while ensuring write exclusivity and fresh reads. The stable machine does not track dirty pages, so it has fewer states (Table~\ref{table:state:parasite}). Pages begin in \t{Exclusive}. Sharing a page with a worker downgrades it to \t{Shared}, while granting write access to a worker downgrades it to \t{Invalid}.

Pages are upgraded on demand. Reading an \t{Invalid} page issues an \t{Unborrow} request to its current writer and upgrades the stable copy to \t{Shared}, while writing a \t{Shared} or \t{Invalid} page issues \t{Unshare} requests and upgrades the stable copy to \t{Exclusive}. To serve worker requests, the manager performs the same upgrades before transferring permissions.

If a worker disappears while a page it holds is \t{Invalid}, the stable machine upgrades its copy to \t{Exclusive} and resumes execution from the latest checkpointed contexts. This is safe: the stable copy already reflects all checkpointed updates, while any newer writes on the worker were never exposed.

\paragraph{Optimistic Execution and Syscalls}\label{sec:mc:syscalls}
Ordinary userspace instructions are executed optimistically: if they fault due to insufficient permissions, \sysname upgrades permissions and retries them. Syscalls are different: if the kernel faults while accessing a user buffer with incompatible permissions, the program may be left in an undefined state, and retrying the syscall would violate semantics.

\sysname therefore ensures that syscalls executed on the stable machine never fault on their user buffers. Before entering the kernel, \sysname locks the corresponding pages in states compatible with the syscall for the entire duration of that syscall. Any incompatible worker request is delayed until the syscall returns. The same principle applies to long-lived kernel registrations: buffers remain locked until a matching deregistration call releases them. To avoid deadlock, worker requests that wait too long are aborted and the offloads at their source are reloaded on the stable machine.

\subsection{Dependency Tracking with MPKs}\label{sec:mc:deps}

Efficient checkpointing requires identifying which threads have interacted, so that only dependent threads are checkpointed together. Dependencies arise through shared memory: if a thread reads data produced by another, then checkpointing the reader also requires checkpointing the writer.

A straightforward way to detect such interactions is to instrument all memory accesses or use standard page protections. Both are too expensive: instrumentation slows every access, while page protections apply process-wide and would fault even when a thread rereads its own data.

\sysname instead uses MPKs to obtain \emph{per-thread} permissions over memory. The key idea is to make the common case free: a thread should access its own working set without overhead, and only the \emph{first} cross-thread access should fault. Intuitively, pages written by a thread are placed in groups that only it may access.
When another thread later touches such a page, the fault reveals a dependency; \sysname records it and extends permissions via a PKRU update~(\S\ref{sec:background:mpk}), allowing later accesses to that group without further faults. If multiple threads write the same page or group, they are conservatively treated as mutually dependent. Since dependencies are tracked per offload interval, a thread loses access to all pages upon offload or checkpoint; later accesses rebuild its permissions and reveal dependencies for the new interval.

Changing a page's MPK group is expensive, so \sysname cannot afford to move pages between groups on every trapped access or checkpoint. Instead, it uses mostly stable groups that capture common access patterns (i.e., past readers and writers). This improves efficiency, but a page's current group no longer precisely reflects recent accessors: treating all associated threads as dependent would be very conservative. \sysname therefore also tracks the groups a page has belonged to since its last checkpoint and combines that list with the permissions actually granted to threads, thereby recovering dependencies less conservatively without frequent group changes. Algorithm~\ref{lst:mpknotting} shows this logic.

\begin{figure}[t]
\begin{lstlisting}[caption={Conservative dependency tracking on faults},label={lst:mpknotting}]
upon fault(thread t, page p, access rw):
    curr_group = maybe_update_group(p, t, rw)
    grant_access(t, curr_group, rw)   // PKRU update
    G = groups_since_last_checkpoint(p)
    R, W = {}, {}
    for g in G:
        R @$\cup$=@ threads_with_read_access(g)
        W @$\cup$=@ threads_with_write_access(g)
    record_dependencies(R, W)
    resume_execution(t)
\end{lstlisting}
\end{figure}

In the work-stealing scenario of Section~\ref{sec:overview:example}, when $T_2$ first accesses a queue previously written by $T_1$, the access faults, revealing a dependency. \sysname records this dependency and extends $T_2$'s permissions so it can access the queue. Subsequent queue accesses incur no further cost, as permissions and dependencies have already been established. Eventually, checkpointing $T_2$ also checkpoints $T_1$, as shown in Figure~\ref{fig:mpk-example}.

\subsection{Per-Thread Dirty Tracking with MPKs}\label{sec:mc:dirty}

Thread-level checkpointing requires identifying which pages the jointly checkpointed threads modified since their last checkpoint. \sysname builds on Linux's soft-dirty mechanism~\cite{emelyanov-softdirty-2013}, which tracks writes by clearing page write permissions and resetting a per-page ``dirty'' bit so that later writes fault and the kernel marks the corresponding pages dirty again. However, the stock mechanism operates at process granularity: clearing dirty bits affects all pages, requires scanning all memory, and typically forces stopping all threads to avoid missing concurrent writes, making it unsuitable for frequent fine-grained checkpointing.

\sysname extends this mechanism to operate at the granularity of MPK groups. As discussed in Section~\ref{sec:mc:deps}, MPKs isolate the pages a thread may access and modify. When a set of threads is checkpointed, those threads are paused, and other threads do not have write access to their groups (and no new access is granted). \sysname can therefore safely scan just those groups, report the pages dirtied in the current interval, and clear their dirty bits so writes are tracked again for the next checkpoint, without pausing unrelated threads.

This is implemented with a small kernel extension to the existing soft-dirty scan. Since that scan already iterates over all VMAs---contiguous memory ranges that notably have uniform MPK groups---\sysname mainly adds a group-aware filter: for each VMA, the extension checks whether its group belongs to the jointly checkpointed threads, skips it otherwise, and clears dirty bits if it does. It then returns exactly the pages whose bits were cleared.

As a result, \sysname processes only the pages relevant to the checkpoint without blocking other threads. For example, when $T_1$ is checkpointed (\S\ref{sec:overview:example}), only pages modified by $T_1$ are reported, while concurrent writes by $T_2$ are ignored.

\subsection{Deferred Protection with MPKs}\label{sec:mc:protect}

TMCP frequently needs to restrict accesses for coherence. For example, when a page is downgraded (e.g., to \t{Invalid}), later accesses that violate its state must fault. Enforcing every such transition with standard page protection is expensive: \t{mprotect} incurs kernel crossings and TLB invalidations.

\sysname instead observes that many downgrades are short-lived and predictable. When a thread is offloaded, pages such as its stack or working set are often transferred to a worker and then upgraded again when the thread reloads. During that interval, the thread is not running locally, so no protection is needed against it. \sysname therefore defers the protection whenever current MPK permissions already prevent other threads from violating the downgrade, and upgrades the page again before its usual accessors resume.

Algorithm~\ref{lst:mpkaution} shows the resulting logic, which applies on both the stable machine and workers using the MPK groups introduced in Section~\ref{sec:mc:deps}. Upon a page's downgrade, \sysname first checks whether current PKRU permissions already rule out accesses that would violate the new state. If so, the protection is deferred. It is applied only if a later PKRU update would make such accesses possible (e.g., when a thread resumes). If the page is upgraded first, the deferred protection is simply discarded. \sysname also prefetches pages and the permissions they will need after reload, so many deferred protections are canceled before they are ever applied.

\begin{figure}[t]
\begin{lstlisting}[caption={Deferred page protection},label={lst:mpkaution}]
upon downgrade(page p, state s):
    if safe_under_pkru(p, s): defer(p, s)
    else: apply_protection(p, s)

upon pkru_update(thread t):
    for each page p with deferred protections:
        if pkru_allows_violation(t, p, state(p)):
            apply_protection(p, state(p))

upon upgrade(page p, state s):
    cancel_deferred(p, s)
\end{lstlisting}
\end{figure}

For example, when $T_1$ is offloaded, its stack is transferred to the worker and downgraded on the stable machine. Because $T_1$ is not running locally and other threads lack access to its stack under MPKs, the protection can be deferred. When $T_1$ later reloads, its stack is prefetched and upgraded, so the deferred protection is canceled without ever being applied.

Standard page protections are still needed when current PKRU permissions do not already make the downgrade safe. For example, $T_C$ may update on the stable machine a page that the worker can currently read through a larger group accessible to $T_1$ and $T_2$. If that page must be downgraded while the rest of the group remains accessible, some thread could still read it under its current group permissions. In that case, \sysname applies \t{mprotect} immediately.

%% file: content/implementation.tex
\section{Implementation}\label{sec:impl}

Our prototype targets x86-64 Linux and consists of 9,229 lines of C++17. Most of \sysname is implemented in userspace, including offloading, memory coherence, and MPK-based dependency tracking. The only kernel component is a small extension to improve soft-dirty handling with groups (\S\ref{sec:mc:dirty}).

\paragraph{Hooking compute-heavy sections.}
\sysname is agnostic to how offload points are identified. In our prototype, we intercept calls to selected dynamic-library functions, motivated by the observation that many compute-intensive tasks rely on specialized libraries (e.g., ML frameworks, codecs, and numerical libraries~\cite{ffmpeg,pytorch,tensorflow,numpy-programming-2020,virtanen2020scipy}). For each such function, a wrapper takes precedence at load time and initiates offloading before invoking the original function.

\paragraph{Handling faults and MPKs.}
\sysname relies on page fault signals to track dependencies and keep memory coherent. The manager installs signal handlers in the application, and workers install the same handlers in each thread prepared to host offloads. x86 MPKs provide only 16 groups per process, so \sysname dynamically merges and recycles groups to stay within this limit. When needed, groups covering the fewest pages are merged, which may reduce precision but remains safe as it conservatively increases dependencies.

\paragraph{Tracking memory layout.}\label{sec:impl:shadow}
\sysname maintains a copy of VMAs~\cite{midguard-virtmem,prefetched-address-translation,translation-ranger} to track which pages may be offloaded and with which rights. The manager initializes this state from \t{/proc/self/maps} and mirrors subsequent mapping changes (e.g., \t{mmap}, \t{mprotect}, \t{brk}). It also detects mappings that cannot be safely offloaded (e.g., memory shared with other processes or device mappings) and forces reloads on access. To avoid conflicts with worker-private mappings, workers are restarted if necessary to obtain non-overlapping layouts.

\paragraph{Intercepting syscalls.}
We intercept all syscalls with negligible overhead using zpoline~\cite{zpoline}. This is used to trigger reloads or para-offloads on workers (\S\ref{sec:overview:example}, \S\ref{sec:para}) and to lock user buffers and track VMAs on the stable machine (\S\ref{sec:mc:syscalls}).

\paragraph{Isolating the manager.}
The manager runs in-process to control the application's flow and resources without costly inter-process synchronization. To avoid interference, it is loaded with \t{dlmopen} and \t{LM\_ID\_NEWLM}~\cite{glibc-dlmopen}, giving it a separate namespace and private copies of dependencies (e.g., \t{libc}). As a result, only application code is instrumented and offloaded.

%% file: content/evaluation.tex
\section{Evaluation}\label{sec:evaluation}

This section answers five questions:
\begin{enumerate}[leftmargin=*]
\item When does \sysname lower deployment cost (\S\ref{sec:eval:e2e})?
\item What is the cost of offloading and checkpointing (\S\ref{sec:eval:lat})?
\item How much does para-offloading amortize this cost (\S\ref{sec:eval:para})?
\item What is the impact of worker reclamations (\S\ref{sec:eval:reclamations})?
\item How much does each MPK mechanism contribute (\S\ref{sec:eval:ablations})?
\end{enumerate}

Our testbed comprises AWS \t{c6i} VMs configured as per Table~\ref{table:hwspecs}. Experiments run in Docker containers with CPU and memory limits matching that of cloud VMs~\cite{ec2-spot}: compute ranges from 1x to 16x two-way hyper-threaded cores and memory from 4 to 64\,GiB. All containers can page to EBS~\cite{ebs}.

\begin{table}[t]
\smallskip
\setlength{\tabcolsep}{2pt}
\caption{Configuration details of our \t{c6i.8xlarge} VMs.}
\label{table:hwspecs}
\small
\centering
\begin{tabular}{cl}
\toprule
\textbf{vCPU}     & 8c/16t Intel Platinum 8375C @ 2.90--3.50\,GHz \\
\textbf{Memory}   & 64\,GiB DDR4 @ 3200\,MT/s \\
\textbf{Storage}  & 10\,Gbps 64\,GiB EBS \\
\textbf{Network}  & 12.5\,Gbps / Cluster placement (ping $\approx$65\,\us) \\
\textbf{Software} & Ubuntu 24.04.3 LTS / Linux 6.14.0-1011-aws \\
\bottomrule
\end{tabular}
\end{table}

\subsection{End-to-End Benefits}\label{sec:eval:e2e}

This subsection asks when \sysname improves the cost efficiency of complete applications.

\paragraph{Methodology.}
Each \sysname deployment uses two containers on distinct machines: a weak stable machine and a stronger transient worker. All reported costs include both machines. We compare against two baselines: (1) a purely on-demand deployment, which provides the same guarantees as \sysname but uses only expensive resources; and (2) a purely transient deployment, which is cheaper but does not tolerate reclamations. 
Because transient prices vary across providers, regions, and time, we report throughput per dollar assuming a representative 60\% transient discount~\cite{azure-spot-pricing,gcp-spot,ec2-spot}. This assumption is illustrative: changing the discount rescales the cost curves, but does not change the underlying throughput trends.

We evaluate six applications spanning different compute-memory mixes, each built around widely used libraries. These includes image classification using PyTorch/ResNet50~\cite{pytorch,resnet50}, speech transcription using Whisper.cpp~\cite{whisper-cpp}, cryptographic workloads based on secp256k1~\cite{secp256k1}, and scientific computing using OneMKL~\cite{onemkl,black-scholes-book}. We also include two data-serving applications backed by SQLite~\cite{sqlite} and \t{std::map}~\cite{cpp-map}.

\subsubsection{Compute-heavy workloads}\label{sec:eval:e2e:computeheavy}

We first ask whether \sysname lowers the cost of compute-heavy workloads.

\begin{figure}
 \includegraphics[width=\columnwidth]{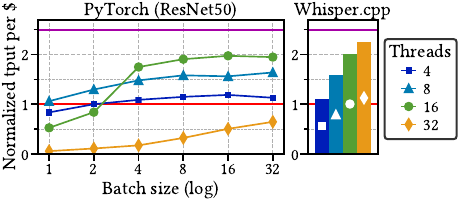}
 \caption{Normalized throughput per dollar of ML inference workloads. The red line ($y{=}1$) marks parity with a purely on-demand deployment; the purple line ($y{=}2.5$) is the upper bound given by a fully transient deployment under a 60\% discount assumption.}
 \label{fig:ml-app}
\end{figure}

\begin{figure}
 \includegraphics[width=\columnwidth]{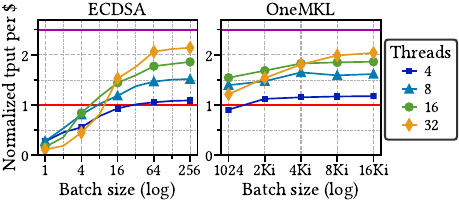}
 \caption{Normalized throughput per dollar of pure-compute workloads, with the same interpretation as Figure~\ref{fig:ml-app}.}
 \label{fig:compute-app}
\end{figure}

For ML inference, we run 4--32 parallel instances either natively or via \sysname. Native deployments use just enough cores to run all instances in parallel. \sysname uses a 2-core stable machine and a worker with enough cores to host all offloads. PyTorch classifies batches of 1--32 images; Whisper.cpp transcribes 10-second samples. Offloading starts at inference and ends at batch or sample completion.

Figure~\ref{fig:ml-app} shows that 4-thread \sysname deployments offer only marginal gains: the stable machine cost is still a large fraction of the total deployment cost. As parallelism grows, this fixed cost is amortized better. For PyTorch, 8- and 16-thread deployments improve throughput per dollar by up to 64\% and 98\%, respectively. At 32 threads, however, the stable machine becomes a bottleneck: starting at 16 threads and striking at 32, memory pressure causes swapping during checkpoint writeback. Whisper.cpp writes back much less memory per offload than PyTorch (24\,KiB vs. 38\,MiB), so it scales better: 8-, 16-, and 32-thread deployments improve throughput per dollar by 58\%, 101\%, and 125\%, respectively.

Overall, \sysname can roughly halve the cost of inference workloads with modest write sets.
Yet, the contrast between PyTorch and Whisper.cpp highlights a key limitation of application-oblivious checkpointing: scratchpad memory is checkpointed even when it will never be read again. Supporting mechanisms to identify such scratchpad regions could further reduce checkpointing overhead.

\vspace{1em}

We next consider pure-compute workloads with little memory pressure. secp256k1 signs and verifies batches of 1--256 signatures; OneMKL prices 2,000 options using 1Ki--16Ki Monte Carlo samples.

Figure~\ref{fig:compute-app} shows the same overall trend. For batches of 256 signatures, 4-, 8-, 16-, and 32-thread \sysname deployments improve throughput per dollar by 9\%, 52\%, 86\%, and 114\%, respectively. For OneMKL with 16Ki samples, the gains are 18\%, 62\%, 87\%, and 103\%. These workloads also show that useful offloads need not be long: ECDSA generation and verification take only 32\,\us and 42\,\us, yet batches of 8 messages (about 0.6\,ms of work) already amortize the offload cost.

\subsubsection{Memory-heavy workloads}

We next ask whether \sysname is useful when transient resources mainly provide additional memory rather than compute.

\begin{figure}
 \includegraphics[width=\columnwidth]{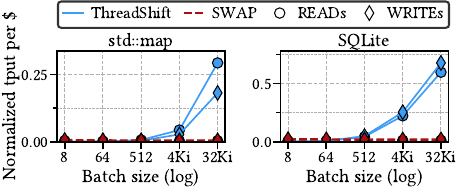}
 \caption{Normalized throughput per dollar of memory-heavy workloads relative to an on-demand deployment that keeps the full dataset in RAM.}
 \label{fig:memory-app}
\end{figure}

We evaluate SQLite and \t{std::map} on a 12\,GiB dataset with read and write batches from 1 to 32Ki operations. We compare three deployments: (1) a pure on-demand deployment with enough RAM to hold all data, (2) a 4\,GiB on-demand deployment forced to swap, and (3) \sysname combining a 4\,GiB on-demand machine with a 16\,GiB transient worker. A warm-up ensures the worker acquires the full dataset.

Figure~\ref{fig:memory-app} shows that neither swapping nor \sysname beats keeping the dataset in on-demand RAM. For \t{std::map}, swap-based deployments achieve only 0.4\% of the normalized throughput of the RAM-resident baseline, while \sysname reaches 18\% for writes and 29\% for reads. For SQLite, swap reaches 0.1\% and 0.2\% for 32Ki writes and reads, while \sysname reaches 60\% and 68\%. \sysname outperforms swapping because lookups run against RAM on the worker, but it remains worse than a fully RAM-resident deployment: checkpointing forces modified data back to the stable machine, effectively reintroducing swapping.

The takeaway is that \sysname is primarily a compute offloader: merely offloading memory is not enough.

\subsection{Offloading and Checkpointing Latency}\label{sec:eval:lat}

This subsection asks how much offloading costs in end-to-end latency, and how much faster \sysname checkpoints than coarse-grained process checkpointing.

We use a synthetic workload in which one offloaded thread modifies 1--4Ki pages before reloading, and measure the end-to-end latency of the offload/reload cycle. To model contention, we also run 0, 3, or 7 independent background offloads on the same worker; each background thread repeatedly writes 1--4Ki pages. We repeat each configuration 1,000 times and report medians.

\begin{figure}
 \includegraphics[width=\columnwidth]{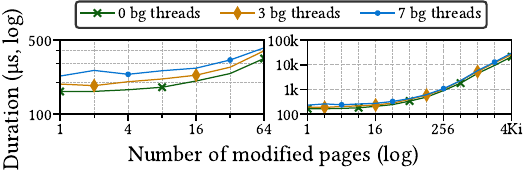}
 \caption{End-to-end offload latency for a thread that writes 1--4Ki pages, with 0, 3, or 7 independent background offloads.}
 \label{fig:offload}
\end{figure}

Figure~\ref{fig:offload} shows that the base latency of an offload that writes one page is 164\,\us. Of this, about 65\,\us comes from network latency, about 10\,\us from checkpoint creation, and about 30\,\us from local synchronization. With 3 or 7 background offloads, the base latency rises modestly to 192\,\us and 229\,\us, mostly due to microarchitectural contention. Above 256 modified pages—when the working set exceeds cache capacity—latency grows roughly linearly with the write set and reaches about 30\,ms at 4Ki pages.

\vspace{1em}

We then isolate checkpointing cost and compare against CRIU, the de facto process-level checkpointing system for Linux~\cite{criu}. We do so by running the same workload natively and triggering a CRIU checkpoint whenever \sysname would reload. For fairness, CRIU uses incremental checkpointing and checkpoints to RAM.

\begin{figure}
 \includegraphics[width=\columnwidth]{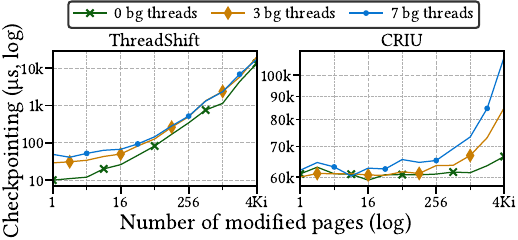}
 \caption{Checkpoint latency for \sysname and CRIU after a thread writes 1--4Ki pages, with 0, 3, or 7 independent background threads running concurrently.}
 \label{fig:checkpoint}
\end{figure}

Figure~\ref{fig:checkpoint} shows that \sysname checkpoints in 10--49\,\us for a one-page write set, and in 13--20\,ms for 4Ki pages. Crucially, this depends only on the pages written by the checkpointed thread, not on the size of the whole process.

CRIU is over three orders of magnitude slower on small checkpoints, and remains substantially slower on large write sets. Its base latency is 61\,ms because it checkpoints the full process state, including kernel objects. Moreover, CRIU cannot distinguish writes by unrelated threads, so its latency grows with both page count and thread count, reaching 67, 84, and 110\,ms when 1, 4, and 8 threads each write 4Ki pages. It also pauses all background threads, whereas \sysname pauses only the threads being checkpointed.

Thus, \sysname's thread granularity and partial offloading reduce checkpoint latency from tens of milliseconds to tens of microseconds on small write sets.

\subsection{Network Para-Offloading}\label{sec:eval:para}

This subsection asks whether para-offloading enables networked servers to benefit from \sysname despite short per-request compute times.

We build a server that replies to 128\,B requests with 128\,B replies after a configurable processing delay. We deploy it (1) natively, (2) with plain \sysname offloading, and (3) with \sysname plus network para-offloading. In the last case, requests are forwarded directly to the worker and replies are batched; we vary the batch size from 2 to 128. A client with 1\,ms ping sends 8Ki requests, from which we measure throughput and latency.

\begin{figure}
 \includegraphics[width=\columnwidth]{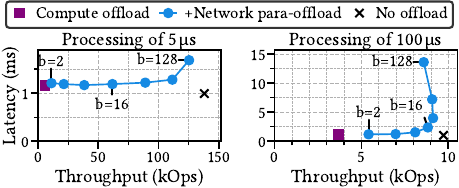}
 \caption{Latency/throughput tradeoff for a server deployed natively, with plain offloading, or with network para-offloading.}
 \label{fig:network}
\end{figure}

Figure~\ref{fig:network} shows that for 5\,\us requests, native throughput is 138\,kOps at 1\,ms latency. Plain offloading almost preserves latency but drops throughput to 5.7\,kOps, as each request incurs a full offload/reload cycle. Batching restores throughput: batches of 64 replies reach 112\,kOps at 1.28\,ms.

For 100\,\us requests, native throughput is 9.7\,kOps, while plain offloading reaches 3.7\,kOps. Here the base offload cost is better amortized, and batching only needs to be modest: batches of 8 replies reach 8.1\,kOps with 1.5\,ms latency.

Thus, network para-offloading brings \sysname close to native per-core throughput for server workloads, at the cost of a modest latency increase.

\subsection{Impact of Reclamations}\label{sec:eval:reclamations}

We now evaluate the performance impact of reclamations.

We measure the throughput of the secp256k1 workload while reclaiming its worker. We use batch sizes of 128, 512, and 4Ki, which determine how often the offloaded thread returns and therefore bound the amount of work that may be lost. We consider two cases: sudden reclamations, which lose all work since the last checkpoint, and notified reclamations, which allow threads to return before the worker disappears. After reclamation, execution continues briefly on the stable machine before being re-offloaded to a fresh worker.

\begin{figure}
 \includegraphics[width=\columnwidth]{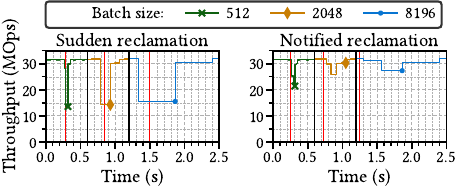}
 \caption{Throughput during sudden and notified worker reclamations for three offload durations.}
 \label{fig:fault}
\end{figure}

Figure~\ref{fig:fault} shows that sudden reclamations cause a throughput dip proportional to the amount of in-flight work. For batch size 512, the dip is narrow, indicating that most of the cost is recomputation rather than failover. With notifications, no work is lost, but throughput still dips briefly while threads return and a new worker is initialized. The cold-start penalty lasts for roughly two offloads, during which VMAs, code, and MPK groups are re-established.

Overall, reclamations cause only a brief throughput dip, mainly proportional to the amount of lost work. Notifications reduce this overhead, but are not required for fast recovery.

\subsection{Mechanism Ablations}\label{sec:eval:ablations}

This subsection investigates how much each MPK-based mechanism contributes in isolation.

\paragraph{Dependency tracking.}
We first disable the MPK-based dependency tracker while keeping the rest of \sysname unchanged. We use a synthetic application with one long-running offloaded thread and 1, 3, or 7 short-running ones. Short threads repeatedly write 16--4Ki pages in batches of 1,000 iterations, while the long-running thread writes 16 pages per batch but runs for 10M iterations. Without dependency tracking, every checkpoint of a short thread at the end of a batch also checkpoints the long-running one.

\begin{figure}
 \includegraphics[width=\columnwidth]{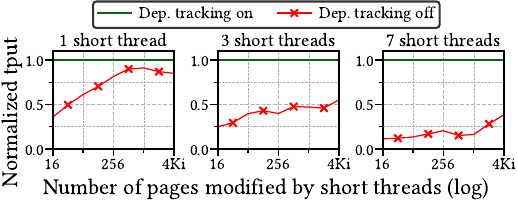}
 \caption{Throughput of the long-running offload with dependency tracking disabled, normalized to the full system.}
 \label{fig:mpkatching}
\end{figure}

Figure~\ref{fig:mpkatching} shows that unnecessary joint checkpoints are very costly. When short threads write only 16 pages, the throughput of the long-running thread drops by 64\%, 76\%, and 90\% with 1, 3, and 7 short threads, respectively. Dependency tracking is therefore essential when threads have mismatched checkpoint frequencies.

\paragraph{Dirty tracking.}
We next disable the per-group soft-dirty kernel extension while keeping dependency tracking enabled, using the same workload as above.

\begin{figure}
 \includegraphics[width=\columnwidth]{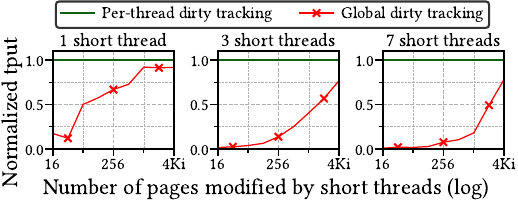}
 \caption{Throughput of the long-running offload with per-thread dirty tracking disabled, normalized to the full system.}
 \label{fig:mpklearing}
\end{figure}

Figure~\ref{fig:mpklearing} shows a similar but stronger degradation. Without per-group dirty tracking, reading and clearing soft-dirty bits requires stopping the entire application to avoid missing concurrent writes. As a result, even threads unrelated to the checkpoint are repeatedly paused. Under frequent checkpointing, these global pauses accumulate and significantly reduce throughput. Per-thread dirty tracking restricts dirty-bit handling to the relevant threads and is therefore critical to preserve concurrency during frequent checkpointing.

\paragraph{Deferred protection.}
Finally, we evaluate deferred protection on a synthetic workload with one offloaded thread that repeatedly writes a configurable number of pages. At each offload, ownership of the thread's stack moves to the worker and then back to the stable machine on reload. We compare the full mechanism against variants with only prefetching (anticipating page upgrades), only deferred protections (delaying them until allowed PKRU accesses would violate permissions), and neither optimization.

\begin{figure}
 \includegraphics[width=\columnwidth]{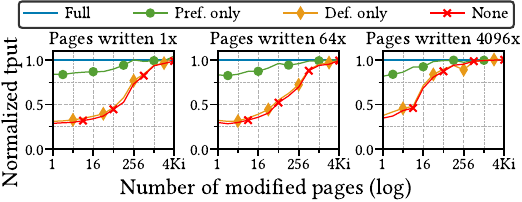}
 \caption{Normalized throughput with deferred protection fully enabled, partially enabled, or disabled.}
 \label{fig:mpkaution}
\end{figure}

Figure~\ref{fig:mpkaution} shows that deferred protection matters primarily for short offloads. When pages are touched only once, disabling the mechanism reduces throughput by up to 73\%. Prefetching alone already doubles throughput by avoiding many faults and coherence round trips. In contrast, deferred protections alone bring little benefit: when the thread reloads, pages such as its stack remain in a downgraded state with deferred protections, which become unsafe and must therefore be enforced. As a result, the protection cost is merely shifted in time rather than avoided, and remains on the critical path. Combining both techniques avoids most protection syscalls and yields a further gain of up to 19\%.

\vspace{1em}
These ablations show a clear division of labor: dependency tracking avoids unnecessary joint checkpoints, per-thread dirty tracking prevents unrelated threads from being stalled, and deferred protection and prefetching remove much of the coherence overhead of short offloads.

%% file: content/discussion.tex
\section{Discussion, Limitations \& Extensions}
\label{sec:discussion}

\paragraph{When is \sysname effective?}
\sysname is most effective for applications whose hot paths are compute-heavy, have modest write sets, and amortize offloading over repeated work. It is less effective for memory-bound applications or workloads that modify large data, as checkpointing must write these changes back to the stable machine.

\paragraph{The role of MPKs.}
MPKs help make \sysname practical for multi-threaded applications. They enable \sysname to track dependencies efficiently, track writes without global synchronization, and defer memory protections. Without MPKs, these mechanisms would be much more expensive.

\paragraph{Automatically identifying offload points.}
Our prototype offloads at selected shared-library calls, which capture many compute-heavy regions in practice. Automatically identifying profitable offload points would improve usability, but appears challenging in practice.

\paragraph{Leveraging faster data paths.}
\sysname's efficiency depends on data movement latency. Faster networking, e.g., RDMA, could reduce offload and checkpoint costs. However, integrating direct memory access with protections is challenging due to concurrency and coherence constraints.

\paragraph{Signals and vDSOs.}
Our prototype does not support signals and requires disabling vDSOs. Signals introduce asynchronous control flow that complicates migration and checkpointing, while vDSOs rely on invariants that do not hold across machines. Supporting both would improve compatibility but requires careful handling.

%% file: content/related.tex
\section{Related Work} \label{sec:related}

\sysname relates to prior work on transient computing, checkpointing and migration, compute offloading, and Memory Protection Keys (MPKs).

\paragraph{Reclamation-Tolerant Applications.}
Prior work studied how to design applications for transient resources, including data stores~\cite{spot-mem-cache,elastic-datastore-spot,blend-spot-memstore}, ML training systems~\cite{bamboo,varuna,DBLP:conf/hotcloud/WagenlanderMLP20}, web services~\cite{spotweb,serverless-spot}, and batch processors~\cite{spoton, DBLP:journals/cluster/LinMF12}. These approaches rely on statelessness or recomputation~\cite{bamboo,varuna,spotweb,serverless-spot,spoton}, or use replication~\cite{spot-mem-cache,elastic-datastore-spot,blend-spot-memstore}. More generally, crash-tolerant applications may still fail under frequent, correlated reclamations~\cite{t-basir}, which follow a different fault model. Frameworks like Kappa~\cite{kappa} structure applications to support checkpointing, but require rewriting. In contrast, \sysname operates transparently on unmodified applications.

\paragraph{Transparent Checkpointing \& Migration.}
Checkpointing and migration have been widely studied for VMs~\cite{postcopy-mig-prepag-selfbal,cloudnet,vm-handoff-edge,snowflock,hexo}, processes~\cite{survey-rollback, cruz, zap, zhong2001crak, sprite, accent, locus-1, amoeba, demos-migration, charlotte, chorus, v-system-1, criu, condor, responsiv-repl-container}, containers~\cite{criu-containers,criu-efficient-containers,process-mig-for-linux-2006}, and managed runtimes~\cite{java-bytecode-thrd-mig1, java-bytecode-thrd-mig2, jessica2}. Recent work extends these techniques to transient environments~\cite{hotspot-server-hop, spotcheck, agg-vm-borrow-ressources, DBLP:conf/eurosys/JiaSSRW16}, but assume sufficient time to migrate. \sysname instead tolerates abrupt reclamations without such assumptions.

Like prior systems, \sysname uses incremental checkpointing~\cite{survey-rollback}. Existing approaches detect modified pages via instrumentation~\cite{lmc}, hardware dirty bits~\cite{ldt}, or OS support such as CRIU’s soft-dirty mechanism~\cite{criu, emelyanov-softdirty-2013}. \sysname extends the latter to thread granularity using MPKs, enabling fast checkpointing without pausing unrelated threads.

\paragraph{Compute Offloading.}
Several systems support partial offloading of computation. Some target managed runtimes such as the JVM~\cite{comet, meteor, dist-jvm-thread-mig, jessica2}, while others provide frameworks for writing offloadable applications~\cite{thinkair, maui, clonecloud, js-offload, sdcf-webassembly-offload, quicksand}. Distributed shared-memory systems~\cite{millipede,thread-mig-in-dsm,ariadne,arachne} provide a shared-memory abstraction across machines but require application changes. Rapid~\cite{rapid-gpgpu-offload} focuses on GPU offloading. In contrast, \sysname offloads threads of unmodified native applications with transparent memory coherence.

\paragraph{Memory Protection Keys.}
MPKs have primarily been used for isolation and security, including execute-only memory~\cite{xom-switch}, synchronization debugging~\cite{kard}, memory monitoring~\cite{fgmambompk}, and library sandboxing~\cite{threadlock}, as well as language-level isolation~\cite{enclosure, pkrusafe}. Other work uses MPKs for fast in-process isolation~\cite{uprocess, endokernel} or intra-kernel protection~\cite{iskios}. 
In contrast, \sysname uses MPKs not for security, but to track inter-thread dependencies, identify per-thread memory modifications, and reduce the cost of enforcing coherence.

Several systems address MPK limitations, such as key virtualization~\cite{libmpk, epk, vdom} and hardware extensions~\cite{specmpk}. These approaches are complementary to \sysname.

%% file: content/conclusion.tex
\section{Conclusion} \label{sec:conclusion}

We presented \sysname, the first system that transparently offloads threads of unmodified Linux applications to transient resources while preserving correctness under reclamations. By operating at thread granularity, \sysname enables fast checkpoints and offloads only compute-heavy tasks.

The main challenge is to checkpoint individual threads despite shared memory.
\sysname addresses it with three novel MPK-based mechanisms that track inter-thread dependencies, identify per-thread memory modifications, and reduce the cost of enforcing coherence across machines.

Our prototype shows that these mechanisms make transparent thread-level offloading practical on commodity clouds. \sysname offloads in as little as 164\,\us, checkpoints up to three orders of magnitude faster than CRIU, and reduces deployment cost by up to 56\% vs on-demand deployments.

%% file: references.bib
@string{MICRO="International Symposium on Microarchitecture (MICRO)"}

@string{SOSP="ACM Symposium on Operating Systems Principles (SOSP)"}

@string{ASPLOS="International Conference on Architectural Support for Programming Languages and Operating Systems (ASPLOS)"}

@string{HPCA = "IEEE Symposium on High Performance Computer Architecture (HPCA)"}

@string{ATC = "USENIX Annual Technical Conference (ATC)"}

@string{SoCC = "Symposium on Cloud Computing (SoCC)"}

@string{HPDC = "Symposium on High-Performance Parallel and Distributed Computing (HPDC)"}

@string{NSDI = "USENIX Symposium on Networked Systems Design and Implementation (NSDI)"}

@string{DSN = "International Conference on Dependable Systems and Networks (DSN)"}

@string{OSDI = "USENIX Symposium on Operating System Design and Implementation (OSDI)"}

@string{SOSP = "ACM Symposium on Operating Systems Principles (SOSP)"}

@string{EuroSys = "European Conference on Computer Systems (EuroSys)"}

@misc{use-azure-spot,
  title        = {Use Azure Spot Virtual Machines},
  author       = {Microsoft},
  year         = 2024,
  url          = {https://learn.microsoft.com/en-us/azure/virtual-machines/spot-vms},
  note         = {Accessed: December 6, 2024}
}

@misc{azure-spot-pricing,
  title        = {Azure Spot Virtual Machines pricing},
  author       = {Microsoft},
  year         = 2024,
  url          = {https://azure.microsoft.com/en-us/pricing/spot-advisor/},
  note         = {Accessed: December 6, 2024}
}

@misc{gcp-spot,
  title        = {Google Cloud Documentation: Spot VMs},
  author       = {Google},
  year         = 2024,
  url          = {https://cloud.google.com/compute/docs/instances/spot},
  note         = {Accessed: December 6, 2024}
}

@misc{ec2-spot,
  title        = {Amazon EC2 Spot Instances},
  author       = {Amazon Web Services},
  year         = 2024,
  url          = {https://aws.amazon.com/ec2/spot/},
  note         = {Accessed: December 6, 2024}
}

@misc{onemkl,
  title        = {Intel® oneAPI Math Kernel Library (oneMKL)},
  author       = {Intel Corporation},
  year         = 2024,
  url          = {https://www.intel.com/content/www/us/en/developer/tools/oneapi/onemkl.html},
  note         = {Accessed: December 6, 2024}
}

@misc{sqlite,
  title        = {{SQLite}},
  author       = {Hipp, Richard and other {SQLite contributors}},
  year         = 2024,
  url          = {https://www.sqlite.org/index.html},
  note         = {Accessed: December 6, 2024}
}

@misc{cpp-map,
title        = {std::map - C++ Reference},
author       = {The cppreference contributors},
year         = 2024,
url          = {https://en.cppreference.com/w/cpp/container/map},
note         = {Accessed: December 6, 2024}
}

@misc{whisper-cpp,
  title        = {whisper.cpp: Port of OpenAI's Whisper model in C/C++},
  author       = {Gerganov, Georgi and other {whisper.cpp contributors}},
  year         = 2024,
  url          = {https://github.com/ggerganov/whisper.cpp}
}

@misc{secp256k1,
  title        = {Optimized C library for EC operations on curve secp256k1},
  author       = {Pieter Wuille and contributors},
  year         = 2013,
  url          = {https://github.com/bitcoin-core/secp256k1}
}

@misc{tensorflow,
  author = {Martín Abadi and Ashish Agarwal and Paul Barham and Eugene Brevdo and Zhifeng Chen and Craig Citro and Greg S. Corrado and Andy Davis and Jeffrey Dean and Matts Dusenberry and Sanjay Ghemawat and Ian Goodfellow and Andrew Harp and Geoffrey Irving and Michael Isard and Yangqing Jia and Rafal Jozefowicz and Lukasz Kaiser and Menglong Zhu and Rajat Monga and Sherry Moore and Derek Murray and Chris Olah and David Petrov and Pieter Abbeel and Ruslan Salakhutdinov and Ilya Sutskever and Kunal Talwar and Paul Tucker and Vijay Vasudevan and Fernanda Viegas and Martin Wattenberg and Martin Wicke and Yuan Yu and Xiaoqiang Zheng},
  title = {{TensorFlow}: Large-Scale Machine Learning on Heterogeneous Systems},
  year = {2015},
  url = {https://www.tensorflow.org/}
}

@inproceedings{jessica2,
author = {Zhu, Wenzhang and Wang, Cho-Li and Lau, Francis C. M.},
title = {JESSICA2: A Distributed Java Virtual Machine with Transparent Thread Migration Support},
year = {2002},
isbn = {0769517455},
publisher = {IEEE Computer Society},
address = {USA},
booktitle = {Proceedings of the IEEE International Conference on Cluster Computing},
pages = {381},
series = {CLUSTER '02},
doi = {10.5555/792762.793331}
}

@inproceedings{java-bytecode-thrd-mig1,
author = {Truyen, Eddy and Robben, Bert and Vanhaute, Bart and Coninx, Tim and Joosen, Wouter and Verbaeten, Pierre},
title = {Portable Support for Transparent Thread Migration in Java},
year = {2000},
isbn = {354041052X},
publisher = {Springer-Verlag},
address = {Berlin, Heidelberg},
booktitle = {Proceedings of the Second International Symposium on Agent Systems and Applications and Fourth International Symposium on Mobile Agents},
pages = {29–43},
numpages = {15},
series = {ASA/MA 2000},
doi = {10.5555/647629.732581}
}

@inproceedings{java-bytecode-thrd-mig2,
author = {Sakamoto, Takahiro and Sekiguchi, Tatsurou and Yonezawa, Akinori},
title = {Bytecode Transformation for Portable Thread Migration in Java},
year = {2000},
isbn = {354041052X},
publisher = {Springer-Verlag},
address = {Berlin, Heidelberg},
booktitle = {Proceedings of the Second International Symposium on Agent Systems and Applications and Fourth International Symposium on Mobile Agents},
pages = {16–28},
numpages = {13},
series = {ASA/MA 2000},
doi = {10.5555/647629.732576}
}

@INPROCEEDINGS{dist-jvm-thread-mig,
  author={Wenzhang Zhu and Cho-Li Wang and Lau, F.C.M.},
  booktitle={2003 International Conference on Parallel Processing, 2003. Proceedings.}, 
  title={Lightweight transparent Java thread migration for distributed JVM}, 
  year={2003},
  volume={},
  number={},
  pages={465-472},
  doi={10.1109/ICPP.2003.1240611}}

@inproceedings{comet,
author = {Gordon, Mark S. and Jamshidi, D. Anoushe and Mahlke, Scott and Mao, Z. Morley and Chen, Xu},
title = {COMET: code offload by migrating execution transparently},
year = {2012},
isbn = {9781931971966},
publisher = {USENIX Association},
address = {USA},
booktitle = {Proceedings of the 10th USENIX Conference on Operating Systems Design and Implementation},
pages = {93–106},
numpages = {14},
location = {Hollywood, CA, USA},
series = {OSDI'12},
doi={10.5555/2387880.2387890}
}

@inproceedings{clonecloud,
author = {Chun, Byung-Gon and Ihm, Sunghwan and Maniatis, Petros and Naik, Mayur and Patti, Ashwin},
title = {CloneCloud: elastic execution between mobile device and cloud},
year = {2011},
isbn = {9781450306348},
publisher = {Association for Computing Machinery},
address = {New York, NY, USA},
url = {https://doi.org/10.1145/1966445.1966473},
doi = {10.1145/1966445.1966473},
booktitle = {Proceedings of the Sixth Conference on Computer Systems},
pages = {301–314},
numpages = {14},
location = {Salzburg, Austria},
series = {EuroSys '11}
}

@INPROCEEDINGS{thinkair,
  author={Kosta, Sokol and Aucinas, Andrius and Pan Hui and Mortier, Richard and Xinwen Zhang},
  booktitle={2012 Proceedings IEEE INFOCOM}, 
  title={ThinkAir: Dynamic resource allocation and parallel execution in the cloud for mobile code offloading}, 
  year={2012},
  volume={},
  number={},
  pages={945-953},
  doi={10.1109/INFCOM.2012.6195845}}

@INPROCEEDINGS{meteor,
  author={Jaemin Lee and Yuhun Jun and Euiseong Seo},
  booktitle={2017 IEEE International Conference on Pervasive Computing and Communications (PerCom)}, 
  title={An enhanced DSM model for computation offloading}, 
  year={2017},
  volume={},
  number={},
  pages={69-78},
  doi={10.1109/PERCOM.2017.7917852},
  organization={IEEE}}

@inproceedings{maui,
author = {Cuervo, Eduardo and Balasubramanian, Aruna and Cho, Dae-ki and Wolman, Alec and Saroiu, Stefan and Chandra, Ranveer and Bahl, Paramvir},
title = {MAUI: making smartphones last longer with code offload},
year = {2010},
isbn = {9781605589855},
publisher = {Association for Computing Machinery},
address = {New York, NY, USA},
url = {https://doi.org/10.1145/1814433.1814441},
doi = {10.1145/1814433.1814441},
booktitle = {Proceedings of the 8th International Conference on Mobile Systems, Applications, and Services},
pages = {49–62},
numpages = {14},
location = {San Francisco, California, USA},
series = {MobiSys '10}
}

@article{js-offload,
author = {Park, Sehoon and Chen, Qichen and Han, Hyuck and Yeom, Heon Y.},
title = {Design and evaluation of mobile offloading system for web-centric devices},
year = {2014},
issue_date = {April 2014},
publisher = {Academic Press Ltd.},
address = {GBR},
volume = {40},
number = {C},
issn = {1084-8045},
journal = {J. Netw. Comput. Appl.},
month = apr,
pages = {105–115},
numpages = {11},
doi = {10.5555/2773807.2774033}
}

@article{rapid-gpgpu-offload,
author = {Montella, Raffaele and Kosta, Sokol and Oro, David and Vera, Javier and Fernández, Carles and Palmieri, Carlo and Di Luccio, Diana and Giunta, Giulio and Lapegna, Marco and Laccetti, Giuliano},
title = {Accelerating Linux and Android applications on low-power devices through remote GPGPU offloading},
journal = {Concurrency and Computation: Practice and Experience},
volume = {29},
number = {24},
doi = {https://doi.org/10.1002/cpe.4286},
url = {https://onlinelibrary.wiley.com/doi/abs/10.1002/cpe.4286},
eprint = {https://onlinelibrary.wiley.com/doi/pdf/10.1002/cpe.4286},
year = {2017}
}

@article{sdcf-webassembly-offload,
author = {Nithya, S. and Sangeetha, M. and Prethi, K. N. Apinaya and Sahoo, Kshira Sagar and Panda, Sanjaya Kumar and Gandomi, Amir H.},
title = {SDCF: A Software-Defined Cyber Foraging Framework for Cloudlet Environment},
year = {2020},
issue_date = {Dec. 2020},
publisher = {IEEE Press},
volume = {17},
number = {4},
issn = {1932-4537},
url = {https://doi.org/10.1109/TNSM.2020.3015657},
doi = {10.1109/TNSM.2020.3015657},
journal = {IEEE Trans. on Netw. and Serv. Manag.},
month = {dec},
pages = {2423–2435},
numpages = {13}
}

@inproceedings{vm-handoff-edge,
author = {Ha, Kiryong and Abe, Yoshihisa and Eiszler, Thomas and Chen, Zhuo and Hu, Wenlu and Amos, Brandon and Upadhyaya, Rohit and Pillai, Padmanabhan and Satyanarayanan, Mahadev},
title = {You can teach elephants to dance: agile VM handoff for edge computing},
year = {2017},
isbn = {9781450350877},
publisher = {Association for Computing Machinery},
address = {New York, NY, USA},
url = {https://doi.org/10.1145/3132211.3134453},
doi = {10.1145/3132211.3134453},
booktitle = {Proceedings of the Second ACM/IEEE Symposium on Edge Computing},
articleno = {12},
numpages = {14},
location = {San Jose, California},
series = {SEC '17}
}

@inproceedings{lmc,
author = {Vogt, Dirk and Giuffrida, Cristiano and Bos, Herbert and Tanenbaum, Andrew S.},
title = {Lightweight Memory Checkpointing},
year = {2015},
isbn = {9781479986293},
publisher = {IEEE Computer Society},
address = {USA},
url = {https://doi.org/10.1109/DSN.2015.45},
doi = {10.1109/DSN.2015.45},
booktitle = {Proceedings of the 2015 45th Annual IEEE/IFIP International Conference on Dependable Systems and Networks},
pages = {474–484},
numpages = {11},
series = {DSN '15}
}

@inproceedings{remus,
author = {Cully, Brendan and Lefebvre, Geoffrey and Meyer, Dutch and Feeley, Mike and Hutchinson, Norm and Warfield, Andrew},
title = {Remus: high availability via asynchronous virtual machine replication},
year = {2008},
isbn = {1119995555221},
publisher = {USENIX Association},
address = {USA},
booktitle = {Proceedings of the 5th USENIX Symposium on Networked Systems Design and Implementation},
pages = {161–174},
numpages = {14},
location = {San Francisco, California},
series = {NSDI'08},
doi={10.5555/1387589.1387601}
}

@inproceedings{plover,
author = {Wang, Cheng and Chen, Xusheng and Jia, Weiwei and Li, Boxuan and Qiu, Haoran and Zhao, Shixiong and Cui, Heming},
title = {Plover: fast, multi-core scalable virtual machine fault-tolerance},
year = {2018},
isbn = {9781931971430},
publisher = {USENIX Association},
address = {USA},
booktitle = {Proceedings of the 15th USENIX Conference on Networked Systems Design and Implementation},
pages = {483–499},
numpages = {17},
location = {Renton, WA, USA},
series = {NSDI'18},
doi = {10.5555/3307441.3307483}
}

@inproceedings{hexo,
author = {Olivier, Pierre and Mehrab, A. K. M. Fazla and Lankes, Stefan and Karaoui, Mohamed Lamine and Lyerly, Robert and Ravindran, Binoy},
title = {HEXO: Offloading HPC Compute-Intensive Workloads on Low-Cost, Low-Power Embedded Systems},
year = {2019},
isbn = {9781450366700},
publisher = {Association for Computing Machinery},
address = {New York, NY, USA},
url = {https://doi.org/10.1145/3307681.3325408},
doi = {10.1145/3307681.3325408},
booktitle = {Proceedings of the 28th International Symposium on High-Performance Parallel and Distributed Computing},
pages = {85–96},
numpages = {12},
location = {Phoenix, AZ, USA},
series = {HPDC '19}
}

@article{cloudnet,
author = {Wood, Timothy and Ramakrishnan, K. K. and Shenoy, Prashant and van der Merwe, Jacobus},
title = {CloudNet: dynamic pooling of cloud resources by live WAN migration of virtual machines},
year = {2011},
issue_date = {July 2011},
publisher = {Association for Computing Machinery},
address = {New York, NY, USA},
volume = {46},
number = {7},
issn = {0362-1340},
url = {https://doi.org/10.1145/2007477.1952699},
doi = {10.1145/2007477.1952699},
journal = {SIGPLAN Not.},
month = {mar},
pages = {121–132},
numpages = {12}
}

@inproceedings{snowflock,
author = {Lagar-Cavilla, Horacio Andr\'{e}s and Whitney, Joseph Andrew and Scannell, Adin Matthew and Patchin, Philip and Rumble, Stephen M. and de Lara, Eyal and Brudno, Michael and Satyanarayanan, Mahadev},
title = {SnowFlock: rapid virtual machine cloning for cloud computing},
year = {2009},
isbn = {9781605584829},
publisher = {Association for Computing Machinery},
address = {New York, NY, USA},
url = {https://doi.org/10.1145/1519065.1519067},
doi = {10.1145/1519065.1519067},
booktitle = {Proceedings of the 4th ACM European Conference on Computer Systems},
pages = {1–12},
numpages = {12},
location = {Nuremberg, Germany},
series = {EuroSys '09}
}

@inproceedings{postcopy-mig-prepag-selfbal,
author = {Hines, Michael R. and Gopalan, Kartik},
title = {Post-copy based live virtual machine migration using adaptive pre-paging and dynamic self-ballooning},
year = {2009},
isbn = {9781605583754},
publisher = {Association for Computing Machinery},
address = {New York, NY, USA},
url = {https://doi.org/10.1145/1508293.1508301},
doi = {10.1145/1508293.1508301},
booktitle = {Proceedings of the 2009 ACM SIGPLAN/SIGOPS International Conference on Virtual Execution Environments},
pages = {51–60},
numpages = {10},
location = {Washington, DC, USA},
series = {VEE '09}
}

@inproceedings {responsiv-repl-container,
author = {Diyu Zhou and Yuval Tamir},
title = {{RRC}: Responsive Replicated Containers},
booktitle = {2022 USENIX Annual Technical Conference (USENIX ATC 22)},
year = {2022},
isbn = {978-1-939133-29-61},
address = {Carlsbad, CA},
pages = {85--100},
url = {https://www.usenix.org/conference/atc22/presentation/zhou-diyu},
publisher = {USENIX Association},
month = jul
}

@article{ariadne,
  title={Ariadne: Architecture of a portable threads system supporting thread migration},
  author={Mascarenhas, Edward and Rego, Vernon},
  journal={Software: Practice and Experience},
  volume={26},
  number={3},
  pages={327--356},
  year={1996},
  publisher={Wiley Online Library}
}

@article{arachne,
author = {Dimitrov, Bozhidar and Rego, Vernon},
title = {Arachne: A Portable Threads System Supporting Migrant Threads on Heterogeneous Network Farms},
year = {1998},
issue_date = {May 1998},
publisher = {IEEE Press},
volume = {9},
number = {5},
issn = {1045-9219},
url = {https://doi.org/10.1109/71.679216},
doi = {10.1109/71.679216},
journal = {IEEE Trans. Parallel Distrib. Syst.},
month = {may},
pages = {459–469},
numpages = {11}
}

@misc{criu,
  title        = {{{CRIU}: Checkpoint/Restore In Userspace}},
  author       = {{CRIU} Development Team},
  year         = 2024,
  url          = {https://criu.org/},
  note         = {Accessed: December 6, 2024}
}

@InProceedings{criu-efficient-containers,
author="Stoyanov, Radostin
and Kollingbaum, Martin J.",
editor="Yokota, Rio
and Weiland, Mich{\`e}le
and Shalf, John
and Alam, Sadaf",
title="Efficient Live Migration of Linux Containers",
booktitle="High Performance Computing",
year="2018",
publisher="Springer International Publishing",
address="Cham",
pages="184--193",
isbn="978-3-030-02465-9",
doi={10.1007/978-3-030-02465-9_13}
}

@InProceedings{criu-containers,
author="Pickartz, Simon
and Eiling, Niklas
and Lankes, Stefan
and Razik, Lukas
and Monti, Antonello",
editor="Taufer, Michela
and Mohr, Bernd
and Kunkel, Julian M.",
title="Migrating LinuX Containers Using CRIU",
booktitle="High Performance Computing",
year="2016",
publisher="Springer International Publishing",
address="Cham",
pages="674--684",
isbn="978-3-319-46079-6"
}

@inproceedings{process-mig-for-linux-2006,
  title={Design and implementation of a process migration system for the Linux environment},
  author={Vasudevan, Nalini and Venkatesh, Prasanna},
  booktitle={3rd International Conference on Neural, Parallel and Scientific Computations},
  pages={1--8},
  year={2006}
}

@ARTICLE{thread-mig-in-dsm,
  author={Thitikamol, K. and Keleher, P.},
  journal={Proceedings of the IEEE}, 
  title={Thread migration and communication minimization in DSM systems}, 
  year={1999},
  volume={87},
  number={3},
  pages={487-497},
  doi={10.1109/5.747869}}

@article{millipede,
author = {Itzkovitz, Ayal and Schuster, Assaf and Shalev, Lea},
title = {Thread migration and its applications in distributed shared memory systems},
year = {1998},
issue_date = {July 1, 1998},
publisher = {Elsevier Science Inc.},
address = {USA},
volume = {42},
number = {1},
issn = {0164-1212},
url = {https://doi.org/10.1016/S0164-1212(98)00008-9},
doi = {10.1016/S0164-1212(98)00008-9},
journal = {J. Syst. Softw.},
month = {jul},
pages = {71–87},
numpages = {17}
}

@inproceedings{blend-spot-memstore,
author = {Xu, Zichen and Stewart, Christopher and Deng, Nan and Wang, Xiaorui},
title = {Blending on-demand and spot instances to lower costs for in-memory storage},
year = {2016},
publisher = {IEEE Press},
url = {https://doi.org/10.1109/INFOCOM.2016.7524348},
doi = {10.1109/INFOCOM.2016.7524348},
booktitle = {IEEE INFOCOM 2016 - The 35th Annual IEEE International Conference on Computer Communications},
pages = {1–9},
numpages = {9},
location = {San Francisco, CA, USA}
}

@article{elastic-datastore-spot,
  title={An Elastic Ephemeral Datastore using Cheap, Transient Cloud Resources},
  author={Brodmann, Malte and Ioannou, Nikolas and Metzler, Bernard and Pfefferle, Jonas and Klimovic, Ana},
  journal={arXiv preprint arXiv:2205.11261},
  year={2022},
  doi={10.48550/arXiv.2205.11261}
}

@inproceedings{serverless-spot,
author = {Zhang, Yanqi and Goiri, \'{I}\~{n}igo and Chaudhry, Gohar Irfan and Fonseca, Rodrigo and Elnikety, Sameh and Delimitrou, Christina and Bianchini, Ricardo},
title = {Faster and Cheaper Serverless Computing on Harvested Resources},
year = {2021},
isbn = {9781450387095},
publisher = {Association for Computing Machinery},
address = {New York, NY, USA},
url = {https://doi.org/10.1145/3477132.3483580},
doi = {10.1145/3477132.3483580},
booktitle = {Proceedings of the ACM SIGOPS 28th Symposium on Operating Systems Principles},
pages = {724–739},
numpages = {16},
location = {Virtual Event, Germany},
series = {SOSP '21}
}

@inproceedings{varuna,
author = {Athlur, Sanjith and Saran, Nitika and Sivathanu, Muthian and Ramjee, Ramachandran and Kwatra, Nipun},
title = {Varuna: scalable, low-cost training of massive deep learning models},
year = {2022},
isbn = {9781450391627},
publisher = {Association for Computing Machinery},
address = {New York, NY, USA},
url = {https://doi.org/10.1145/3492321.3519584},
doi = {10.1145/3492321.3519584},
booktitle = {Proceedings of the Seventeenth European Conference on Computer Systems},
pages = {472–487},
numpages = {16},
location = {Rennes, France},
series = {EuroSys '22}
}

@inproceedings {bamboo,
author = {John Thorpe and Pengzhan Zhao and Jonathan Eyolfson and Yifan Qiao and Zhihao Jia and Minjia Zhang and Ravi Netravali and Guoqing Harry Xu},
title = {Bamboo: Making Preemptible Instances Resilient for Affordable Training of Large {DNNs}},
booktitle = {20th USENIX Symposium on Networked Systems Design and Implementation (NSDI 23)},
year = {2023},
isbn = {978-1-939133-33-5},
address = {Boston, MA},
pages = {497--513},
url = {https://www.usenix.org/conference/nsdi23/presentation/thorpe},
publisher = {USENIX Association},
month = apr
}

@inproceedings{spotcheck,
author = {Sharma, Prateek and Lee, Stephen and Guo, Tian and Irwin, David and Shenoy, Prashant},
title = {SpotCheck: designing a derivative IaaS cloud on the spot market},
year = {2015},
isbn = {9781450332385},
publisher = {Association for Computing Machinery},
address = {New York, NY, USA},
url = {https://doi.org/10.1145/2741948.2741953},
doi = {10.1145/2741948.2741953},
booktitle = {Proceedings of the Tenth European Conference on Computer Systems},
articleno = {16},
numpages = {15},
location = {Bordeaux, France},
series = {EuroSys '15}
}

@inproceedings{spot-mem-cache,
author = {Wang, Cheng and Urgaonkar, Bhuvan and Gupta, Aayush and Kesidis, George and Liang, Qianlin},
title = {Exploiting Spot and Burstable Instances for Improving the Cost-efficacy of In-Memory Caches on the Public Cloud},
year = {2017},
isbn = {9781450349383},
publisher = {Association for Computing Machinery},
address = {New York, NY, USA},
url = {https://doi.org/10.1145/3064176.3064220},
doi = {10.1145/3064176.3064220},
booktitle = {Proceedings of the Twelfth European Conference on Computer Systems},
pages = {620–634},
numpages = {15},
location = {Belgrade, Serbia},
series = {EuroSys '17}
}

@inproceedings{spoton,
author = {Subramanya, Supreeth and Guo, Tian and Sharma, Prateek and Irwin, David and Shenoy, Prashant},
title = {SpotOn: a batch computing service for the spot market},
year = {2015},
isbn = {9781450336512},
publisher = {Association for Computing Machinery},
address = {New York, NY, USA},
url = {https://doi.org/10.1145/2806777.2806851},
doi = {10.1145/2806777.2806851},
booktitle = {Proceedings of the Sixth ACM Symposium on Cloud Computing},
pages = {329–341},
numpages = {13},
location = {Kohala Coast, Hawaii},
series = {SoCC '15}
}

@inproceedings{hotspot-server-hop,
author = {Shastri, Supreeth and Irwin, David},
title = {HotSpot: automated server hopping in cloud spot markets},
year = {2017},
isbn = {9781450350280},
publisher = {Association for Computing Machinery},
address = {New York, NY, USA},
url = {https://doi.org/10.1145/3127479.3132017},
doi = {10.1145/3127479.3132017},
booktitle = {Proceedings of the 2017 Symposium on Cloud Computing},
pages = {493–505},
numpages = {13},
location = {Santa Clara, California},
series = {SoCC '17}
}

@inproceedings{spotweb,
author = {Ali-Eldin, Ahmed and Westin, Jonathan and Wang, Bin and Sharma, Prateek and Shenoy, Prashant},
title = {SpotWeb: Running Latency-sensitive Distributed Web Services on Transient Cloud Servers},
year = {2019},
isbn = {9781450366700},
publisher = {Association for Computing Machinery},
address = {New York, NY, USA},
url = {https://doi.org/10.1145/3307681.3325397},
doi = {10.1145/3307681.3325397},
booktitle = {Proceedings of the 28th International Symposium on High-Performance Parallel and Distributed Computing},
pages = {1–12},
numpages = {12},
location = {Phoenix, AZ, USA},
series = {HPDC '19}
}

@inproceedings{agg-vm-borrow-ressources,
author = {Chuang, Ho-Ren and Manaouil, Karim and Xing, Tong and Barbalace, Antonio and Olivier, Pierre and Heerekar, Balvansh and Ravindran, Binoy},
title = {Aggregate VM: Why Reduce or Evict VM's Resources When You Can Borrow Them From Other Nodes?},
year = {2023},
isbn = {9781450394871},
publisher = {Association for Computing Machinery},
address = {New York, NY, USA},
url = {https://doi.org/10.1145/3552326.3587452},
doi = {10.1145/3552326.3587452},
booktitle = {Proceedings of the Eighteenth European Conference on Computer Systems},
pages = {469–487},
numpages = {19},
location = {Rome, Italy},
series = {EuroSys '23}
}

@ARTICLE{t-basir,
  author={Alourani, Abdullah and Kshemkalyani, Ajay D. and Grechanik, Mark},
  journal={IEEE Transactions on Parallel and Distributed Systems}, 
  title={T-BASIR: Finding Shutdown Bugs for Cloud-Based Applications in Cloud Spot Markets}, 
  year={2020},
  volume={31},
  number={8},
  pages={1912-1924},
  doi={10.1109/TPDS.2020.2980265}
}

@InProceedings{resnet50,
  author = {He, Kaiming and Zhang, Xiangyu and Ren, Shaoqing and Sun, Jian},
  title = {Deep Residual Learning for Image Recognition},
  booktitle = {Proceedings of the IEEE Conference on Computer Vision and Pattern Recognition (CVPR)},
  month = {June},
  year = {2016},
  doi={10.1109/CVPR.2016.90}
}

@inbook{pytorch,
author = {Paszke, Adam and Gross, Sam and Massa, Francisco and Lerer, Adam and Bradbury, James and Chanan, Gregory and Killeen, Trevor and Lin, Zeming and Gimelshein, Natalia and Antiga, Luca and Desmaison, Alban and K\"{o}pf, Andreas and Yang, Edward and DeVito, Zach and Raison, Martin and Tejani, Alykhan and Chilamkurthy, Sasank and Steiner, Benoit and Fang, Lu and Bai, Junjie and Chintala, Soumith},
title = {PyTorch: an imperative style, high-performance deep learning library},
year = {2019},
publisher = {Curran Associates Inc.},
address = {Red Hook, NY, USA},
booktitle = {Proceedings of the 33rd International Conference on Neural Information Processing Systems},
articleno = {721},
numpages = {12},
doi = {10.5555/3454287.3455008}
}

@inproceedings{openfoam,
  title={OpenFOAM: A C++ library for complex physics simulations},
  author={Jasak, Hrvoje and Jemcov, Aleksandar and Tukovic, Zeljko and others},
  booktitle={International workshop on coupled methods in numerical dynamics},
  volume={1000},
  pages={1--20},
  year={2007},
  organization={Dubrovnik, Croatia)}
}

@article{lammps,
title = {{LAMMPS} - a flexible simulation tool for particle-based materials modeling at the atomic, meso, and continuum scales},
journal = {Computer Physics Communications},
volume = {271},
pages = {108171},
year = {2022},
issn = {0010-4655},
doi = {https://doi.org/10.1016/j.cpc.2021.108171},
url = {https://www.sciencedirect.com/science/article/pii/S0010465521002836},
publisher={Elsevier},
author = {Aidan P. Thompson and H. Metin Aktulga and Richard Berger and Dan S. Bolintineanu and W. Michael Brown and Paul S. Crozier and Pieter J. {in 't Veld} and Axel Kohlmeyer and Stan G. Moore and Trung Dac Nguyen and Ray Shan and Mark J. Stevens and Julien Tranchida and Christian Trott and Steven J. Plimpton},
}

@article{homomorphic-encrypt-survey,
author = {Acar, Abbas and Aksu, Hidayet and Uluagac, A. Selcuk and Conti, Mauro},
title = {A Survey on Homomorphic Encryption Schemes: Theory and Implementation},
year = {2018},
issue_date = {July 2019},
publisher = {Association for Computing Machinery},
address = {New York, NY, USA},
volume = {51},
number = {4},
issn = {0360-0300},
url = {https://doi.org/10.1145/3214303},
doi = {10.1145/3214303},
journal = {ACM Comput. Surv.},
month = jul,
articleno = {79},
numpages = {35}
}

@inproceedings{argon2,
  author={Biryukov, Alex and Dinu, Daniel and Khovratovich, Dmitry},
  booktitle={2016 IEEE European Symposium on Security and Privacy (EuroS\&P)}, 
  title={Argon2: New Generation of Memory-Hard Functions for Password Hashing and Other Applications}, 
  year={2016},
  volume={},
  number={},
  pages={292-302},
  doi={10.1109/EuroSP.2016.31}
}

@article{zero-knowledge-blockchain,
  author={Sun, Xiaoqiang and Yu, F. Richard and Zhang, Peng and Sun, Zhiwei and Xie, Weixin and Peng, Xiang},
  journal={IEEE Network}, 
  title={A Survey on Zero-Knowledge Proof in Blockchain}, 
  year={2021},
  volume={35},
  number={4},
  pages={198-205},
  doi={10.1109/MNET.011.2000473}
}

@book{black-scholes-book,
  title={The Black--Scholes Model},
  author={Capi{\'n}ski, Marek and Kopp, Ekkehard},
  year={2012},
  publisher={Cambridge University Press},
  place={Cambridge},
  series={Mastering Mathematical Finance},
  collection={Mastering Mathematical Finance},
  doi={10.1017/CBO9781139026130}
}

@manual{intel2019developer,
  author       = {Intel Corporation},
  title        = {Intel® 64 and IA-32 Architectures Software Developer's Manual},
  year         = {2019},
  edition      = {Volume 3: System Programming Guide},
  url          = {https://www.intel.com/content/www/us/en/developer/articles/technical/intel-sdm.html},
  note         = {Accessed: December 6, 2024}
}

@inproceedings{midguard-virtmem,
author = {Gupta, Siddharth and Bhattacharyya, Atri and Oh, Yunho and Bhattacharjee, Abhishek and Falsafi, Babak and Payer, Mathias},
title = {Rebooting virtual memory with midgard},
year = {2021},
isbn = {9781450390866},
publisher = {IEEE Press},
url = {https://doi.org/10.1109/ISCA52012.2021.00047},
doi = {10.1109/ISCA52012.2021.00047},
booktitle = {Proceedings of the 48th Annual International Symposium on Computer Architecture},
pages = {512–525},
numpages = {14},
location = {Virtual Event, Spain},
series = {ISCA '21}
}

@inproceedings{prefetched-address-translation,
author = {Margaritov, Artemiy and Ustiugov, Dmitrii and Bugnion, Edouard and Grot, Boris},
title = {Prefetched Address Translation},
year = {2019},
isbn = {9781450369381},
publisher = {Association for Computing Machinery},
address = {New York, NY, USA},
url = {https://doi.org/10.1145/3352460.3358294},
doi = {10.1145/3352460.3358294},
booktitle = {Proceedings of the 52nd Annual IEEE/ACM International Symposium on Microarchitecture},
pages = {1023–1036},
numpages = {14},
location = {Columbus, OH, USA},
series = {MICRO '52}
}

@inproceedings{translation-ranger,
author = {Yan, Zi and Lustig, Daniel and Nellans, David and Bhattacharjee, Abhishek},
title = {Translation ranger: operating system support for contiguity-aware TLBs},
year = {2019},
isbn = {9781450366694},
publisher = {Association for Computing Machinery},
address = {New York, NY, USA},
url = {https://doi.org/10.1145/3307650.3322223},
doi = {10.1145/3307650.3322223},
booktitle = {Proceedings of the 46th International Symposium on Computer Architecture},
pages = {698–710},
numpages = {13},
location = {Phoenix, Arizona},
series = {ISCA '19}
}

@inproceedings {zpoline,
author = {Kenichi Yasukata and Hajime Tazaki and Pierre-Louis Aublin and Kenta Ishiguro},
title = {zpoline: a system call hook mechanism based on binary rewriting},
booktitle = {2023 USENIX Annual Technical Conference (USENIX ATC 23)},
year = {2023},
isbn = {978-1-939133-35-9},
address = {Boston, MA},
pages = {293--300},
url = {https://www.usenix.org/conference/atc23/presentation/yasukata},
publisher = {USENIX Association},
month = jul
}

@manual{glibc-dlmopen,
  title        = {dlopen(3) — Linux manual page},
  author       = {GNU Project},
  year         = {2024},
  url          = {https://man7.org/linux/man-pages/man3/dlmopen.3.html}
}

@manual{glibc-ld,
  title        = {ld.so(8) — Linux manual page},
  author       = {GNU Project},
  year         = {2025},
  url          = {https://man7.org/linux/man-pages/man8/ld.so.8.html}
}

@manual{ffmpeg,
  title        = {FFmpeg},
  author       = {FFmpeg Developers},
  organization = {FFmpeg},
  year         = {2000--},
  url          = {https://ffmpeg.org/},
  note         = {Accessed: December 6, 2024}
}

@article{numpy-programming-2020,
  title={Array programming with NumPy},
  author={Harris, Charles R and Millman, K Jarrod and Van Der Walt, St{\'e}fan J and Gommers, Ralf and Virtanen, Pauli and Cournapeau, David and Wieser, Eric and Taylor, Julian and Berg, Sebastian and Smith, Nathaniel J and others},
  journal={Nature},
  volume={585},
  number={7825},
  pages={357--362},
  year={2020},
  publisher={Nature Publishing Group UK London},
  doi={10.1038/s41586-020-2649-2}
}

@article{virtanen2020scipy,
  title={SciPy 1.0: fundamental algorithms for scientific computing in Python},
  author={Virtanen, Pauli and Gommers, Ralf and Oliphant, Travis E and Haberland, Matt and Reddy, Tyler and Cournapeau, David and Burovski, Evgeni and Peterson, Pearu and Weckesser, Warren and Bright, Jonathan and others},
  journal={Nature methods},
  volume={17},
  number={3},
  pages={261--272},
  year={2020},
  publisher={Nature Publishing Group},
  doi={10.1038/s41592-019-0686-2},
  issn={1548-7105}
}

@inproceedings{DBLP:conf/hotcloud/WagenlanderMLP20,
   author       = {Marcel Wagenl{\"{a}}nder and
                   Luo Mai and
                   Guo Li and
                   Peter R. Pietzuch},
   editor       = {Amar Phanishayee and
                   Ryan Stutsman},
   title        = {Spotnik: Designing Distributed Machine Learning for 
Transient Cloud
                   Resources},
   booktitle    = {12th {USENIX} Workshop on Hot Topics in Cloud 
Computing, HotCloud
                   2020, July 13-14, 2020},
   publisher    = {{USENIX} Association},
   year         = {2020},
   url          = 
{https://www.usenix.org/conference/hotcloud20/presentation/wagenl\%C3\%A4nder},
   bibsource    = {dblp computer science bibliography, https://dblp.org}
}

@inproceedings{DBLP:conf/eurosys/JiaSSRW16,
   author       = {Qin Jia and
                   Zhiming Shen and
                   Weijia Song and
                   Robbert van Renesse and
                   Hakim Weatherspoon},
   editor       = {Yehia Elkhatib and
                   Mohamed Faten Zhani},
   title        = {Smart spot instances for the supercloud},
   booktitle    = {Proceedings of the 3rd Workshop on CrossCloud 
Infrastructures {\&}
                   Platforms, CrossCloud@EuroSys 2016, London, United 
Kingdom, April
                   18-21, 2016},
   pages        = {5:1--5:6},
   publisher    = {{ACM}},
   year         = {2016},
   url          = {https://doi.org/10.1145/2904111.2904114},
   doi          = {10.1145/2904111.2904114},
   bibsource    = {dblp computer science bibliography, https://dblp.org}
}

@article{DBLP:journals/cluster/LinMF12,
   author       = {Heshan Lin and
                   Xiaosong Ma and
                   Wu{-}chun Feng},
   title        = {Reliable MapReduce computing on opportunistic resources},
   journal      = {Clust. Comput.},
   volume       = {15},
   number       = {2},
   pages        = {145--161},
   year         = {2012},
   url          = {https://doi.org/10.1007/s10586-011-0158-7},
   doi          = {10.1007/S10586-011-0158-7},
   bibsource    = {dblp computer science bibliography, https://dblp.org}
}

@inproceedings{kard,
author = {Ahmad, Adil and Lee, Sangho and Fonseca, Pedro and Lee, Byoungyoung},
title = {Kard: lightweight data race detection with per-thread memory protection},
year = {2021},
isbn = {9781450383172},
publisher = {Association for Computing Machinery},
address = {New York, NY, USA},
url = {https://doi.org/10.1145/3445814.3446727},
doi = {10.1145/3445814.3446727},
booktitle = {Proceedings of the 26th ACM International Conference on Architectural Support for Programming Languages and Operating Systems},
pages = {647–660},
numpages = {14},
location = {Virtual, USA},
series = {ASPLOS '21}
}

@article{fgmambompk,
author = {Wang, Rui-bo and Wu, Zhen-wei and Zhang, Wen-zhe and Wu, Hui-jun and Zhang-Yu, Shu-qing and Lu, Kai},
title = {Fine-grained memory access monitoring based on memory protection keys},
publisher = {Computer Engineering \& Science},
year = {2024},
journal = {Computer Engineering \& Science},
volume = {46},
number = {01},
eid = {21},
pages = {21-27},
url = http://joces.nudt.edu.cn/EN/abstract/article_17812.shtml
}

@inproceedings{xom-switch,
  author    = {Daiping Liu and Mingwei Zhang and Ravi Sahita},
  title     = {eXecutable-Only Memory-Switch (XOM-Switch)},
  booktitle = {Black Hat Asia Briefings},
  year      = {2018},
  publisher = {Black Hat Asia}
}

@inproceedings{threadlock,
author = {Blair, William and Robertson, William and Egele, Manuel},
title = {ThreadLock: Native Principal Isolation Through Memory Protection Keys},
year = {2023},
isbn = {9798400700989},
publisher = {Association for Computing Machinery},
address = {New York, NY, USA},
url = {https://doi.org/10.1145/3579856.3595797},
doi = {10.1145/3579856.3595797},
booktitle = {Proceedings of the 2023 ACM Asia Conference on Computer and Communications Security},
pages = {966–979},
numpages = {14},
location = {Melbourne, VIC, Australia},
series = {ASIA CCS '23}
}

@inproceedings{enclosure,
author = {Ghosn, Adrien and Kogias, Marios and Payer, Mathias and Larus, James R. and Bugnion, Edouard},
title = {Enclosure: language-based restriction of untrusted libraries},
year = {2021},
isbn = {9781450383172},
publisher = {Association for Computing Machinery},
address = {New York, NY, USA},
url = {https://doi.org/10.1145/3445814.3446728},
doi = {10.1145/3445814.3446728},
booktitle = {Proceedings of the 26th ACM International Conference on Architectural Support for Programming Languages and Operating Systems},
pages = {255–267},
numpages = {13},
location = {Virtual, USA},
series = {ASPLOS '21}
}

@inproceedings{pkrusafe,
author = {Kirth, Paul and Dickerson, Mitchel and Crane, Stephen and Larsen, Per and Dabrowski, Adrian and Gens, David and Na, Yeoul and Volckaert, Stijn and Franz, Michael},
title = {PKRU-safe: automatically locking down the heap between safe and unsafe languages},
year = {2022},
isbn = {9781450391627},
publisher = {Association for Computing Machinery},
address = {New York, NY, USA},
url = {https://doi.org/10.1145/3492321.3519582},
doi = {10.1145/3492321.3519582},
booktitle = {Proceedings of the Seventeenth European Conference on Computer Systems},
pages = {132–148},
numpages = {17},
location = {Rennes, France},
series = {EuroSys '22}
}

@inproceedings{endokernel,
author = {Yang, Fangfei and Im, Bumjin and Huang, Weijie and Kaoudis, Kelly and Vahldiek-Oberwagner, Anjo and Tsai, Chia-Che and Dautenhahn, Nathan},
title = {Endokernel: a thread safe monitor for lightweight subprocess isolation},
year = {2024},
isbn = {978-1-939133-44-1},
publisher = {USENIX Association},
address = {USA},
booktitle = {Proceedings of the 33rd USENIX Conference on Security Symposium},
articleno = {9},
numpages = {18},
location = {Philadelphia, PA, USA},
series = {SEC '24}
}

@inproceedings{uprocess,
author = {Lin, Jiazhen and Chen, Youmin and Gao, Shiwei and Lu, Youyou},
title = {Fast Core Scheduling with Userspace Process Abstraction},
year = {2024},
isbn = {9798400712517},
publisher = {Association for Computing Machinery},
address = {New York, NY, USA},
url = {https://doi.org/10.1145/3694715.3695976},
doi = {10.1145/3694715.3695976},
booktitle = {Proceedings of the ACM SIGOPS 30th Symposium on Operating Systems Principles},
pages = {280–295},
numpages = {16},
location = {Austin, TX, USA},
series = {SOSP '24}
}

@inproceedings{iskios,
author = {Gravani, Spyridoula and Hedayati, Mohammad and Criswell, John and Scott, Michael L.},
title = {Fast Intra-kernel Isolation and Security with IskiOS},
year = {2021},
isbn = {9781450390583},
publisher = {Association for Computing Machinery},
address = {New York, NY, USA},
url = {https://doi.org/10.1145/3471621.3471849},
doi = {10.1145/3471621.3471849},
booktitle = {Proceedings of the 24th International Symposium on Research in Attacks, Intrusions and Defenses},
pages = {119–134},
numpages = {16},
location = {San Sebastian, Spain},
series = {RAID '21}
}

@inproceedings{libmpk,
author = {Park, Soyeon and Lee, Sangho and Xu, Wen and Moon, Hyungon and Kim, Taesoo},
title = {Libmpk: software abstraction for intel memory protection keys (intel MPK)},
year = {2019},
isbn = {9781939133038},
publisher = {USENIX Association},
address = {USA},
booktitle = {Proceedings of the 2019 USENIX Conference on Usenix Annual Technical Conference},
pages = {241–254},
numpages = {14},
location = {Renton, WA, USA},
series = {USENIX ATC '19}
}

@inproceedings {epk,
author = {Jinyu Gu and Hao Li and Wentai Li and Yubin Xia and Haibo Chen},
title = {{EPK}: Scalable and Efficient Memory Protection Keys},
booktitle = {2022 USENIX Annual Technical Conference (USENIX ATC 22)},
year = {2022},
isbn = {978-1-939133-29-36},
address = {Carlsbad, CA},
pages = {609--624},
url = {https://www.usenix.org/conference/atc22/presentation/gu-jinyu},
publisher = {USENIX Association},
month = jul
}

@inproceedings{vdom,
author = {Yuan, Ziqi and Hong, Siyu and Chang, Rui and Zhou, Yajin and Shen, Wenbo and Ren, Kui},
title = {VDom: Fast and Unlimited Virtual Domains on Multiple Architectures},
year = {2023},
isbn = {9781450399166},
publisher = {Association for Computing Machinery},
address = {New York, NY, USA},
url = {https://doi.org/10.1145/3575693.3575735},
doi = {10.1145/3575693.3575735},
booktitle = {Proceedings of the 28th ACM International Conference on Architectural Support for Programming Languages and Operating Systems, Volume 2},
pages = {905–919},
numpages = {15},
location = {Vancouver, BC, Canada},
series = {ASPLOS 2023}
}

@inproceedings{specmpk,
  author    = {Debpratim Adak and Huiyang Zhou and Eric Rotenberg and Amro Awad},
  title     = {{SpecMPK}: Efficient In-Process Isolation with Speculative and Secure Permission Update Instruction},
  booktitle = {Proceedings of the IEEE International Symposium on High-Performance Computer Architecture (HPCA)},
  year      = {2025},
  series    = {HPCA '25}
}

@inproceedings {quicksand,
author = {Zhenyuan Ruan and Shihang Li and Kaiyan Fan and Seo Jin Park and Marcos K. Aguilera and Adam Belay and Malte Schwarzkopf},
title = {Quicksand: Harnessing Stranded Datacenter Resources with Granular Computing},
booktitle = {22nd USENIX Symposium on Networked Systems Design and Implementation (NSDI 25)},
year = {2025},
isbn = {978-1-939133-46-5},
address = {Philadelphia, PA},
pages = {147--165},
url = {https://www.usenix.org/conference/nsdi25/presentation/ruan},
publisher = {USENIX Association},
month = apr
}

@article{sprite,
author = {Ousterhout, John K. and Cherenson, Andrew R. and Douglis, Frederick and Nelson, Michael N. and Welch, Brent B.},
title = {The Sprite Network Operating System},
year = {1988},
issue_date = {February 1988},
publisher = {IEEE Computer Society Press},
address = {Washington, DC, USA},
volume = {21},
number = {2},
issn = {0018-9162},
url = {https://doi.org/10.1109/2.16},
doi = {10.1109/2.16},
journal = {Computer},
month = feb,
pages = {23–36},
numpages = {14}
}

@article{v-system-1,
author = {Cheriton, David R. and Zwaenepoel, Willy},
title = {The distributed V kernel and its performance for diskless workstations},
year = {1983},
issue_date = {October 1983},
publisher = {Association for Computing Machinery},
address = {New York, NY, USA},
volume = {17},
number = {5},
issn = {0163-5980},
url = {https://doi.org/10.1145/773379.806621},
doi = {10.1145/773379.806621},
journal = {SIGOPS Oper. Syst. Rev.},
month = oct,
pages = {129–140},
numpages = {12}
}

@book{locus-1,
author = {Popek, Gerald J. and Walker, Bruce J.},
title = {The LOCUS distributed system architecture},
year = {1986},
isbn = {0262161028},
publisher = {Massachusetts Institute of Technology},
address = {USA}
}

@article{demos-migration,
author = {Powell, Michael L. and Miller, Barton P.},
title = {Process migration in DEMOS/MP},
year = {1983},
issue_date = {October 1983},
publisher = {Association for Computing Machinery},
address = {New York, NY, USA},
volume = {17},
number = {5},
issn = {0163-5980},
url = {https://doi.org/10.1145/773379.806619},
doi = {10.1145/773379.806619},
journal = {SIGOPS Oper. Syst. Rev.},
month = oct,
pages = {110–119},
numpages = {10}
}

@article{zap,
author = {Osman, Steven and Subhraveti, Dinesh and Su, Gong and Nieh, Jason},
title = {The design and implementation of Zap: a system for migrating computing environments},
year = {2003},
issue_date = {Winter 2002},
publisher = {Association for Computing Machinery},
address = {New York, NY, USA},
volume = {36},
number = {SI},
issn = {0163-5980},
url = {https://doi.org/10.1145/844128.844162},
doi = {10.1145/844128.844162},
journal = {SIGOPS Oper. Syst. Rev.},
month = dec,
pages = {361–376},
numpages = {16}
}

@article{accent,
author = {Rashid, Richard F. and Robertson, George G.},
title = {Accent: A communication oriented network operating system kernel},
year = {1981},
issue_date = {December 1981},
publisher = {Association for Computing Machinery},
address = {New York, NY, USA},
volume = {15},
number = {5},
issn = {0163-5980},
url = {https://doi.org/10.1145/1067627.806593},
doi = {10.1145/1067627.806593},
journal = {SIGOPS Oper. Syst. Rev.},
month = dec,
pages = {64–75},
numpages = {12}
}

@article{amoeba,
author = {Mullender, Sape J. and van Rossum, Guido and Tanenbaum, Andrew S. and van Renesse, Robbert and van Staveren, Hans},
title = {Amoeba: A Distributed Operating System for the 1990s},
year = {1990},
issue_date = {May 1990},
publisher = {IEEE Computer Society Press},
address = {Washington, DC, USA},
volume = {23},
number = {5},
issn = {0018-9162},
url = {https://doi.org/10.1109/2.53354},
doi = {10.1109/2.53354},
journal = {Computer},
month = may,
pages = {44–53},
numpages = {10}
}

@inproceedings{chorus,
author = {Rozier, Marc},
title = {Chorus},
year = {1992},
isbn = {1880446421},
publisher = {USENIX Association},
address = {USA},
booktitle = {Proceedings of the Workshop on Micro-Kernels and Other Kernel Architectures},
pages = {39–70},
numpages = {32}
}

@techreport{zhong2001crak,
  title={CRAK: Linux checkpoint/restart as a kernel module},
  author={Zhong, Hua and Nieh, Jason},
  year={2001},
  institution={Citeseer}
}

@INPROCEEDINGS{cruz,
  author={Janakiraman, G.J. and Santos, J.R. and Subhraveti, D. and Turner, Y.},
  booktitle={2005 International Conference on Dependable Systems and Networks (DSN'05)}, 
  title={Cruz: Application-Transparent Distributed Checkpoint-Restart on Standard Operating Systems}, 
  year={2005},
  volume={},
  number={},
  pages={260-269},
  doi={10.1109/DSN.2005.33}}

@ARTICLE{charlotte,
  author={Artsy, Y. and Finkel, R.},
  journal={Computer}, 
  title={Designing a process migration facility: the Charlotte experience}, 
  year={1989},
  volume={22},
  number={9},
  pages={47-56},
  doi={10.1109/2.35213}}

@article{survey-rollback,
author = {Elnozahy, E. N. (Mootaz) and Alvisi, Lorenzo and Wang, Yi-Min and Johnson, David B.},
title = {A survey of rollback-recovery protocols in message-passing systems},
year = {2002},
issue_date = {September 2002},
publisher = {Association for Computing Machinery},
address = {New York, NY, USA},
volume = {34},
number = {3},
issn = {0360-0300},
url = {https://doi.org/10.1145/568522.568525},
doi = {10.1145/568522.568525},
journal = {ACM Comput. Surv.},
month = sep,
pages = {375–408},
numpages = {34}
}

@article{message-dependencies,
author = {Chandy, K. Mani and Lamport, Leslie},
title = {Distributed snapshots: determining global states of distributed systems},
year = {1985},
issue_date = {Feb. 1985},
publisher = {Association for Computing Machinery},
address = {New York, NY, USA},
volume = {3},
number = {1},
issn = {0734-2071},
url = {https://doi.org/10.1145/214451.214456},
doi = {10.1145/214451.214456},
journal = {ACM Trans. Comput. Syst.},
month = feb,
pages = {63–75},
numpages = {13}
}

@misc{emelyanov-softdirty-2013,
  author       = {Pavel Emelyanov},
  title        = {Soft-Dirty PTEs},
  howpublished = {The Linux Kernel documentation (v6.1)},
  year         = {2013},
  month        = apr,
  url          = {https://www.kernel.org/doc/html/v6.1/admin-guide/mm/soft-dirty.html},
  note         = {Accessed: 2025-08-27}
}

@article{condor,
author = {Thain, Douglas and Tannenbaum, Todd and Livny, Miron},
title = {Distributed computing in practice: the Condor experience: Research Articles},
year = {2005},
issue_date = {February 2005},
publisher = {John Wiley and Sons Ltd.},
address = {GBR},
volume = {17},
number = {2–4},
issn = {1532-0626},
journal = {Concurr. Comput.: Pract. Exper.},
month = feb,
pages = {323–356},
numpages = {34}
}

@INPROCEEDINGS{ldt,
  author={Singh, Rohit and KP, Arun and Mishra, Debadatta},
  booktitle={2022 IEEE 29th International Conference on High Performance Computing, Data, and Analytics (HiPC)}, 
  title={LDT: Lightweight Dirty Tracking of Memory Pages for x86 Systems}, 
  year={2022},
  volume={},
  number={},
  pages={85-94},
  doi={10.1109/HiPC56025.2022.00023}}

@inproceedings{dthreads,
author = {Liu, Tongping and Curtsinger, Charlie and Berger, Emery D.},
title = {Dthreads: efficient deterministic multithreading},
year = {2011},
isbn = {9781450309776},
publisher = {Association for Computing Machinery},
address = {New York, NY, USA},
url = {https://doi.org/10.1145/2043556.2043587},
doi = {10.1145/2043556.2043587},
booktitle = {Proceedings of the Twenty-Third ACM Symposium on Operating Systems Principles},
pages = {327–336},
numpages = {10},
location = {Cascais, Portugal},
series = {SOSP '11}
}

@misc{ebs,
  author       = {Amazon Web Services, Inc.},
  title        = {Amazon Elastic Block Store (EBS) — High-Performance Block Storage},
  howpublished = {\url{https://aws.amazon.com/ebs/}},
  year         = {2025},
  note         = {Accessed: 2025-09-18}
}

@inproceedings{kappa,
author = {Zhang, Wen and Fang, Vivian and Panda, Aurojit and Shenker, Scott},
title = {Kappa: a programming framework for serverless computing},
year = {2020},
isbn = {9781450381376},
publisher = {Association for Computing Machinery},
address = {New York, NY, USA},
url = {https://doi.org/10.1145/3419111.3421277},
doi = {10.1145/3419111.3421277},
booktitle = {Proceedings of the 11th ACM Symposium on Cloud Computing},
pages = {328–343},
numpages = {16},
location = {Virtual Event, USA},
series = {SoCC '20}
}
